\documentclass[prd,aps,floats,twocolumn,superscriptaddress]{revtex4}

\usepackage[dvips]{graphicx}
\usepackage{amssymb}
\usepackage{amsmath}

\usepackage{color}

\begin{document}

\title{Black-Hole Spin Dependence in the Light Curves of Tidal Disruption Events}

\author{Michael Kesden} \email{mhk10@nyu.edu}

\affiliation{Center for Cosmology and Particle Physics, Department of Physics,
New York University, New York, New York 10003}

\date{May 2012}
                            
\begin{abstract}
A star orbiting a supermassive black hole can be tidally disrupted if the black hole's gravitational tidal field
exceeds the star's self gravity at pericenter.  Some of this stellar tidal debris can become gravitationally
bound to the black hole, leading to a bright electromagnetic flare with bolometric luminosity proportional to
the rate at which material falls back to pericenter.  In the Newtonian limit, this flare will have a light curve that
scales as $t^{-5/3}$ if the tidal debris has a flat distribution in binding energy.  We investigate the time
dependence of the black-hole mass accretion rate when tidal disruption occurs close enough the black hole
that relativistic effects are significant.  We find that for orbits with pericenters comparable to the radius of the
marginally bound circular orbit, relativistic effects can double the peak accretion rate and halve the time it
takes to reach this peak accretion rate.  The accretion rate depends on both the magnitude of the black-hole
spin and its orientation with respect to the stellar orbit; for orbits with a given pericenter radius in
Boyer-Lindquist coordinates, a maximal black-hole spin anti-aligned with the orbital angular momentum
leads to the largest peak accretion rate.
\end{abstract}

\maketitle

\section{Introduction} \label{S:intro}

Active galactic nuclei (AGN) are believed to be powered by accretion onto compact objects with masses
$M > 10^5 M_\odot$ \cite{Hoyle:1963}; such objects will inevitably collapse into supermassive black holes
(SBHs) on short timescales \cite{LyndenBell:1969yx}.  Early work \cite{Hills:1975,Young:1977} conjectured 
that these AGN were fueled by the tidal disruption of stars passing too close to the SBHs, and that SBHs
could grow to their observed masses by accreting debris from such tidal disruption events (TDEs).  Although
SBHs are now believed to grow primarily by accreting gas driven into galactic centers by tidal torques during
galactic mergers \cite{Toomre:1972vt,Barnes:1991zz}, interest in TDEs was renewed when it was realized
that they could power bright flares lasting several years in otherwise quiescent galaxies \cite{Rees:1988bf}.

The Roentgensatellit (ROSAT) observed such a flare in soft X-rays in NGC 5905, a galaxy with no previous
indication of nuclear activity \cite{Bade:1996}.  A systematic survey for X-ray flares within the ROSAT All-Sky
Survey discovered five such events, implying a rate of $9.1 \times 10^{-6}$ galaxy$^{-1}$ yr$^{-1}$
consistent with the predicted rate of TDEs \cite{Donley:2002mp}.  Five additional TDE candidates were
discovered in the X-ray by the XMM-Newton Slew Survey \cite{Esquej:2006ff}.   TDEs emit in the UV and
optical as well; several TDE candidates discovered in the UV by the Galaxy Evolution Explorer (GALEX)
were found to have optical counterparts in the Canada-France-Hawaii Telescope Legacy Survey and
Panoramic Survey Telescope and Rapid Response System (Pan-STARRS) Medium Deep Survey
\cite{Gezari:2006fe,Gezari:2007bw,Gezari:2012sa}.  Two new TDE candidates were identified through a
series of rigorous cuts on the many optical transients found in Stripe 82 of the Sloan Digital Sky Survey
(SDSS); this implies that deeper, higher-cadence surveys such as that which will be undertaken by the
Large Synoptic Survey Telescope (LSST) could find thousands of TDEs per year \cite{vanVelzen:2010jp}.
TDEs can also be observed in hard X-rays if they launch jets pointed towards us; the events Sw J1644+57
\cite{Burrows:2011dn,Levan:2011yr,Bloom:2011xk} and Sw J2058+05 \cite{Cenko:2011ys} discovered by
the Burst Alert Telescope on the {\it Swift} satellite are conjectured to be such relativistic tidal disruption
flares.

One of the most exciting applications of observing TDEs is the possibility of measuring SBH spins.  The
magnitude of a SBH's spin depends sensitively on the manner in which that SBH was assembled.  A
nonspinning SBH can attain a dimensionless spin $a/M \simeq 0.998$ after increasing its mass by a
factor of $\sqrt{6}$ through the coherent accretion of gas in the equatorial plane
\cite{Bardeen:1970,Thorne:1974ve}.  If a SBH grows instead through the chaotic accretion of misaligned
clumps of gas \cite{Moderski:1996,King:2008au,Hopkins:2011gb}, its spin magnitude could be much more
modest ($a/M \lesssim 0.2$).  Studies of SBH spins in the context of hierarchical galaxy formation
indicate that a wide range of distributions are possible, depending on one's assumptions about the relative
contributions of coherent accretion, chaotic accretion, and binary mergers to SBH spin evolution
\cite{Volonteri:2004cf,Berti:2008af,Barausse:2012fy}.  Simulations in which gas is accreted coherently or
chaotically lead to dimensionless spin distributions sharply peaked about unity or zero respectively at all
redshifts.  Real SBHs probably grow through an as yet undetermined combination of coherent and chaotic
accretion, suggesting that almost any distribution of SBH spins is possible.  This considerable theoretical
uncertainty emphasizes the urgent need for greater observational constraints.

Fortunately, there has recently been tremendous progress measuring SBH spins using X-ray spectroscopy.
If an AGN accretion disk is surrounded by a hot, nonthermal corona, a portion of the hard X-rays emitted by
this corona may be reprocessed by the colder disk \cite{Guilbert:1988,Lightman:1988eg}.  This hard X-ray
irradiation will cause the 6.4 keV iron K$\alpha$ line in the disk to fluoresce, with much of the flux in this line
coming from the innermost regions around the SBH \cite{Fabian:1989ej,Laor:1991nc}.  The shape and
variability of the line profile is sensitive to general-relativistic effects including SBH spin.  Observations of
fluorescent iron K$\alpha$ lines with the XMM-Newton and Suzaku satellites have been used to constrain
the spin magnitudes of eight nearby AGN, three of which are near maximal ($a/M \gtrsim 0.98$) while the
remaining five are intermediate ($a/M \simeq 0.7$) \cite{Brenneman:2006hw,Brenneman:2011wz}.

Despite these impressive new results, this approach to measuring SBH spins is inherently limited to highly
accreting AGN whose spins may not reflect the SBH population as a whole.  TDEs provide an opportunity
to measure the spins of a less biased SBH population, and spin estimates based on TDEs should have
different systematic errors than previous approaches.  If a TDE launches jets that extract rotational energy
from the SBH via the Blandford-Znajek mechanism \cite{Blandford:1977ds}, the observed peak X-ray
luminosity of the TDE can be used to constrain the SBH spin magnitude \cite{Lei:2011qg}.   If the jet is
misaligned with the SBH spin, the Lense-Thirring effect \cite{Lense:1918zz} will cause the jet axis to
precess, modulating the observed X-ray emission \cite{Stone:2011mz,Lei:2012}.  Both of these approaches
have been applied to the two {\it Swift} events J1644+57 and J2058+05.  Mean SBH spins may also be
estimated from the observed TDE rate \cite{Beloborodov:1992,Ivanov:2005se,Kesden:2011ee}.

In this paper, we focus on yet another way in which TDEs can be used to constrain SBH spins, through the
observed light curve of individual events.  Several previous authors have investigated relativistic effects
during tidal disruption \cite{Luminet:1985,Laguna:1993,Diener:1997,Ivanov:2003}; Laguna {\it et al.}
\cite{Laguna:1993} provide light curves for TDEs by non-spinning SBHs.  If a star with mass $m_\ast$ and
radius $R_\ast$ is on an initially parabolic orbit with pericenter $r_p$, after tidal disruption roughly half of the
stellar mass will become gravitationally bound to the SBH with specific binding energy
\cite{Rees:1988bf}
\begin{equation} \label{E:Etid}
E_{\rm tid} = \frac{GM}{r_p^2} R_\ast = \beta^2 \frac{Gm_\ast}{R_\ast} \left( \frac{M}{m_\ast} \right)^{1/3}
\end{equation}
where $\beta \equiv r_{\rm tid}/r_p$ is the penetration factor, $r_{\rm tid} = (M/m_\ast)^{1/3} R_\ast$ is
the tidal radius, and the binding energy is defined as the positive energy required to unbind the system.
Newtonian orbits with specific binding energy $E$ have orbital period
\begin{equation} \label{E:binper}
t = 2\pi GM (2E)^{-3/2}~,
\end{equation}
implying that the debris will return to pericenter a time
\begin{eqnarray} \label{E:delay}
t_{\rm tid} &=& \beta^{-3} \frac{\pi}{m_\ast} \left( \frac{MR_{\ast}^3}{2G} \right)^{1/2} \nonumber \\
&=& 0.11~{\rm yr}~\beta^{-3} \left( \frac{M}{10^6 M_\odot} \right)^{1/2} \left( \frac{m_\ast}{M_\odot} \right)^{-1}
\left( \frac{R_\ast}{R_\odot} \right)^{3/2}
\end{eqnarray}
after tidal disruption \cite{Lodato:2010xs}.  Internal shocks during this pericenter passage will cause the
debris to settle into an axisymmetric torus on this timescale.  If the viscosity is high enough to allow
accretion on a similar or shorter timescale, Eq.~(\ref{E:binper}) implies that the TDE luminosity
$L \propto |dE/dt| \propto t^{-5/3}$ provided the energy distribution $dm/dE$ of the tidal debris is constant
\cite{Rees:1988bf,Phinney}.  The light curves of several TDE candidates have been reasonably well fit by
$t^{-5/3}$ power laws in the optical, UV, and X-ray
\cite{Gezari:2007bw,Gezari:2012sa,Burrows:2011dn,Komossa:1999xb}.

The $L \propto t^{-5/3}$ dependence of TDE light curves rests on two assumptions: the constancy of
$dm/dE$ and the validity of the Newtonian relation (\ref{E:binper}) between the specific binding energy and
orbital period.  The first of these assumptions was investigated by Lodato, King, and Pringle 
\cite{Lodato:2008fr} (hereafter LKP09), who found that $dm/dE$ did indeed depend on the adiabatic index
$\gamma$ of the tidally disrupted star.  This dependence on $\gamma$ is reflected in the earliest portions of
the TDE light curve; Gezari {\it et al.} \cite{Gezari:2012sa} showed that the TDE candidate PS1-10jh
discovered by Pan-STARRS and GALEX is more consistent with a fully convective star or degenerate core
with $\gamma = 5/3$ than a solar-type star with $\gamma = 4/3$.

The second assumption was investigated by Haas {\it et al.} \cite{Haas:2012bk}, who performed six
numerical-relativity simulations of the tidal disruption of a white dwarf by a $10^3 M_\odot$
intermediate-mass black hole (IMBH).  The IMBH was nonspinning in the first simulation, and in the
remaining five simulations had a spin with the same magnitude ($a/M = 0.6$) but different orientations with
respect to the orbital angular momentum and initial position of the white dwarf.  They found that the prompt
accretion rate at $t \lesssim 6$~s after tidal disruption could vary by almost two orders of magnitude as a
function of spin direction.  Numerical-relativity simulations are computationally expensive however, and
it is not feasible to perform such simulations for the long timescales $ t_{\rm tid}$ given by
Eq.~(\ref{E:delay}) over which debris continues to fall back to pericenter.  Haas {\it et al.} \cite{Haas:2012bk}
instead estimate the fallback time for each fluid element at the end of their simulations assuming that
subsequent motion is on Keplerian ballistic trajectories \cite{Evans:1989qe}.  They find, similar to
LKP09, that $dm/dt \propto t^{-5/3}$ at sufficiently late times.

In this paper, we use the assumption that tidal debris moves on Kerr geodesics to calculate mass accretion
rates as a function of time throughout the full fallback regime.  Our focus will be on the most massive SBHs
where the tidal radius $r_{\rm tid}$ is comparable to the Schwarzschild radius $r_s$; the extreme mass
ratios ($m_\ast/M \lesssim 10^{-6}$) imply that full numerical relativity is not required leading to a vast
reduction in computational expense.  In Sec.~\ref{S:Newt}, we review the Newtonian calculation of the mass
accretion rate given by LKP09.  We then propose a relativistic generalization of this calculation in
Sec.~\ref{S:rel}.  Using this new relativistic framework, we present the accretion rate as a function of SBH
mass, spin, and initial stellar orbit in Sec.~\ref{S:res}.  We discuss the implications of our results and future
applications of this framework in Sec.~\ref{S:disc}.  For the reader's convenience, we have summarized the
Kerr metric and the use of Fermi normal coordinates in Appendices \ref{A:Kerr} and \ref{A:Fermi}.

\section{Newtonian accretion rate} \label{S:Newt}

A simplified model of stellar tidal disruption in the Newtonian limit is presented in Sec. 2 of LKP09.  This
model is premised on the existence of a clear hierarchy between the specific orbital kinetic energy
$E_{\rm orb}$ of the initial star, the energy $E_{\rm tid}$ gained or lost during tidal disruption, and the
self-binding energy $E_{\rm bin}$ of the star \cite{Evans:1989qe}.  These energy scales are given by\begin{subequations} \label{E:Escales}
\begin{eqnarray}
\label{E:Eorb}
E_{\rm orb} &=& \frac{GM}{r_p}~, \\
\label{E:Etid2}
E_{\rm tid} &=& \frac{GM}{r_{p}^2} R_\ast =  \beta q^{1/3} E_{\rm orb}~, \\
\label{E:Ebin}
E_{\rm bin} &=& \frac{Gm_\ast}{R_\ast} = \beta^{-1} q^{2/3} E_{\rm orb}~,
\end{eqnarray}
\end{subequations}
where $r_p$ is the pericenter and $q \equiv m_\ast/M \ll 1$ is the mass ratio.  This hierarchy implies
that to zeroth order in $q$, the specific binding energy of a fluid element of the tidal debris is the same as that
of the center of mass of the initial star.  At $\mathcal{O}(q^{1/3})$, a fluid element located at $\mathbf{r}$
receives a correction to its specific energy
\begin{equation} \label{E:Ecorr}
\Delta E = \frac{GM}{r_{p}^2} \mathbf{\hat{r}}_p \cdot (\mathbf{r} - \mathbf{r}_p)
\end{equation}
where $\mathbf{r}_p$ points from the black hole to the center of the star at pericenter, and vectors with hats
above them are unit vectors.  This correction is of order $E_{\rm tid}$.  The kinetic energy of the fluid element
in the star's center-of-mass frame is of order $E_{\rm bin}$, as is the gravitational binding energy of the fluid
element to the star.  These can therefore be safely neglected to this order in $q$.

Given this hierarchy of energy scales, Rees \cite{Rees:1988bf} approximated the distribution in specific
energy of the tidal debris of a star on an initially parabolic orbit ($E = 0$) as
\begin{eqnarray} \label{E:flatED}
\frac{dm}{dE} &=& \frac{m_\ast}{2E_{\rm tid}} \quad \quad |E| < E_{\rm tid} \nonumber \\
&=& 0 \quad \quad \quad \quad |E| > E_{\rm tid}~.
\end{eqnarray}
As described in Sec.~\ref{S:intro}, this approximation leads to a mass accretion rate that begins a time
$ t_{\rm tid}$ after tidal disruption given by Eq.~(\ref{E:delay}) and falls of as $t^{-5/3}$ thereafter.

\begin{figure}[t!]
\begin{center}
\includegraphics[width=3.5in]{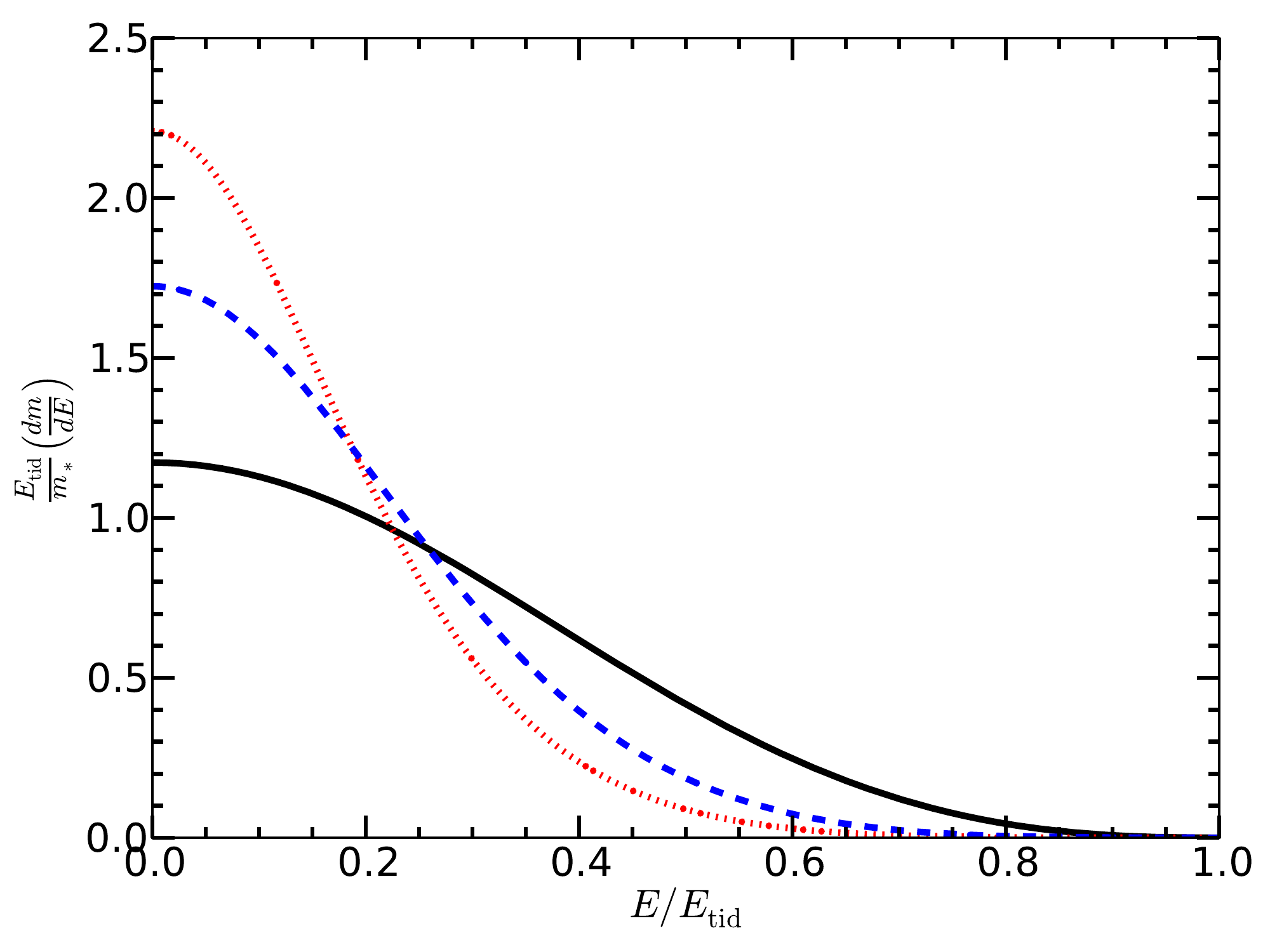}
\end{center}
\caption{The distribution $dm/dE$ of the specific binding energy $E$ for the tidal debris of a star on an
initially parabolic orbit.  The solid black curve corresponds to a adiabatic index $\gamma = 5/3$
appropriate for a fully convective star, while the dashed blue and dotted red curves correspond to
$\gamma = 1.4$ and $4/3$ respectively.}
\label{F:dmdE_Lod}
\end{figure}

LKP09 move beyond this level of approximation, explicitly calculating the
energy distribution of the tidal debris to $\mathcal{O}(q^{1/3})$ for different polytropic equations of state
$P = C\rho^\gamma$.  They begin by solving the Lane-Emden equation to determine the density profile of
the initial star \cite{Shapiro:1983du}
\begin{equation} \label{E:LaEm}
\frac{1}{\xi^2} \frac{d}{d\xi} \left( \xi^2 \frac{d\Theta}{d\xi} \right) = -\Theta^n~,
\end{equation}
where $n \equiv (\gamma -1)^{-1}$ is the polytropic index and $\Theta$ and $\xi$ are dimensionless
variables which give the star's density profile through the relations
\begin{subequations} \label{E:LaEmVar}
\begin{eqnarray}
\label{E:rhoLE}
\rho &=& \rho_c \Theta^n~, \\
\label{E:rLE}
r &=& b\xi~, \\
\label{E:aE}
b &=& \left[ \frac{(n+1)C\rho_{c}^{\gamma-2}}{4\pi G} \right]^{1/2}~,
\end{eqnarray}
\end{subequations}
where $\rho_c$ is the star's central density.  The Lane-Emden equation (\ref{E:LaEm}) can be solved
numerically by beginning at the star's center $\xi = 0$ where $\Theta(0) = 1$, $d\Theta/d\xi(0) = 0$, then
integrating to the star's surface at $\xi_1$ defined by the condition $\Theta(\xi_1) = 0$.  If the specific binding
energy of a fluid element is related to its position by Eq.~(\ref{E:Ecorr}), the energy distribution $dm/dE$ of
the tidal debris is given in terms of the star's initial density profile $\Theta(\xi)$ by
\begin{equation} \label{E:dmdE_Lod}
\frac{E_{\rm tid}}{m_\ast} \frac{dm}{dE} =
\frac{\int_{E/E_{\rm tid}}^1 \Theta^n x dx}{2\int_{0}^1 \Theta^n x^2 dx}~,
\end{equation}
where $x \equiv \xi/\xi_1$ is a rescaled radial coordinate.	  This dimensionless energy distribution for several
adiabatic indices $\gamma$ is shown in Fig.~\ref{F:dmdE_Lod} which reproduces the left panel of Fig.~2 of
LKP09.

\begin{figure}[t!]
\begin{center}
\includegraphics[width=3.5in]{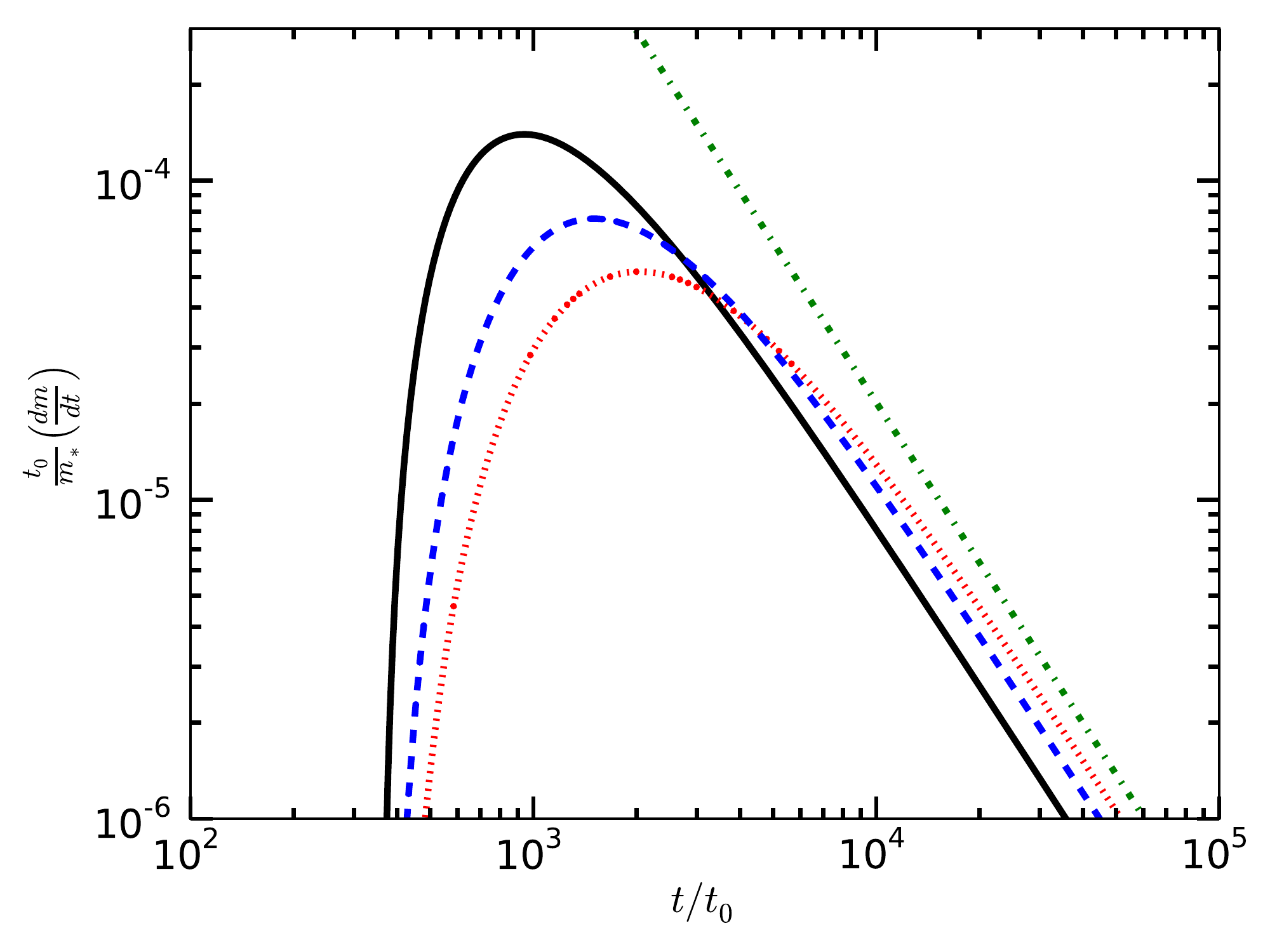}
\end{center}
\caption{The rate $dm/dt$ at which the debris of a tidally disrupted star falls back to pericenter and is
subsequently accreted.  The time $t$ is given in units of $t_0 \equiv 2\pi(r_{p}^3/GM)^{1/2}$, the period of an
orbit with semi-major axis $r_p$.  As in Fig.~\ref{F:dmdE_Lod}, the solid black, dashed blue, and dotted red
curves correspond to adiabatic indices $\gamma = 5/3$, 1.4, and 4/3 respectively.   The dot-dashed green
line shows the canonical $t^{-5/3}$ dependence predicted by \cite{Rees:1988bf,Phinney}.}
\label{F:dmdt_Lod}
\end{figure}

If the orbital period $t$ of a fluid element is given
in terms of its specific binding energy by Eq.~(\ref{E:binper}), the rate $dm/dt$ at which mass falls back to
pericenter and is subsequently accreted is given by
\begin{equation} \label{E:dmdt_Lod}
\frac{t_0}{m_\ast} \frac{dm}{dt} = \frac{1}{3} \frac{r_p}{R_\ast} \left( \frac{E_{\rm tid}}{m_\ast} \frac{dm}{dE}
\right) \left( \frac{t}{t_0} \right)^{-5/3}~,
\end{equation}
where $t_0 \equiv 2\pi(r_{p}^3/GM)^{1/2}$ is the period of an orbit with semi-major axis $r_p$.  We plot this
dimensionless accretion rate for $r_p = 100 R_\ast$ and several values of $\gamma$ in 
Fig.~\ref{F:dmdt_Lod}, a reproduction of the middle panel of Fig.~2 of LKP09.

After introducing this simplified model of tidal disruption in Sec.~2 of their paper, LKP09 devote the rest of
the paper to testing its validity using a series of non-relativistic smoothed-particle-hydrodynamics (SPH)
simulations.  They find that Eq.~(\ref{E:dmdE_Lod}) accurately describes the energy distribution of the tidal
debris after the initial density profile of the star has been homologously expanded by a factor of 1.6 to 2.5
depending on the adiabatic index $\gamma$.  This expansion reflects the stretching of the star by tidal
forces prior to its arrival at pericenter.  The mass accretion rate $dm/dt$ measured in the simulations also 
confirms the predictions shown in Fig.~\ref{F:dmdt_Lod}: a stiffer equation of state (larger $\gamma$)
leads to a higher peak accretion rate and a faster approach to the late-time $t^{-5/3}$ behavior.  The
success of the LKP09 model provides a strong motivation for developing a relativistic version of this model
appropriate to TDEs where the pericenter distance $r_p$ is comparable to the Schwarzschild radius
$r_S = 2GM/c^2$ of the SBH.

\section{Relativistic accretion rate} \label{S:rel}

Kerr geodesics differ from Newtonian orbits in several important respects that need to be incorporated into
our new relativistic model for tidal disruption.  We provide a brief review of Kerr geodesics in
Appendix~\ref{A:Kerr}.  One difference is that Kerr geodesics generally do not close, unlike bound orbits
about a Newtonian point mass.  Since we are interested in the rate at which tidal debris returns to pericenter,
we will focus on the radial period and ignore that the second pericenter will generally occur at different
values of $\theta$ and $\phi$ than the pericenter at which the star was initially disrupted.  This precession
may affect the timescale on which the tidal debris settles into an accretion disk, but this process is beyond
the scope of the current work.

A second difference between Kerr geodesics and Newtonian orbits is that the radial period depends not
only on the specific energy $E$, as in Eq.~(\ref{E:binper}), but also on the specific angular momentum $L_z$
and Carter constant $Q$ as can be seen from Eq.~(\ref{E:drdt}).  Determining the accretion rate $dm/dt$ will
therefore require not just the energy distribution $dm/dE$, but the full joint distribution $d^3m/dEdL_zdQ$ for
all three constants of motion.  Also, as we do not have a nice relation like Eq.~(\ref{E:binper}) for the radial
period, we will have to integrate Eqs.~(\ref{E:EOM}) directly for each fluid element of the tidal debris.

The greatest difference between our relativistic model and the LKP09 model is the manner in which we
determine the distribution $d^3m/dEdL_zdQ$.  In the Newtonian limit, the specific energy
\begin{equation} \label{E:ENewt}
E = c^2 + \frac{1}{2} v^2 + \Phi(r)
\end{equation}
separates into three distinct components: the rest-mass energy $c^2$, the kinetic energy $v^2/2$, and the
gravitational-potential energy $\Phi(r)$.  The rest-mass energy is neglected in the LKP09 model because it
does not affect the orbital motion in the Newtonian limit.  Although the kinetic energy of the center of mass is
equal in magnitude to the gravitational-potential energy for a parabolic orbit, the {\it relative} velocity $v$
between fluid elements at pericenter can be neglected because it leads to energy corrections of order
$E_{\rm bin}$, which according to Eq.~(\ref{E:Escales}) are suppressed by a factor $q^{1/3}/\beta^2$
compared to the dominant corrections of order $E_{\rm tid}$ arising from the gravitational potential.  The
constants of motion $E$, $L_z$, and $Q$ given by Eqs.~(\ref{E:ELz}) and (\ref{E:Qdef}) do not neatly
separate into components that depend only on $r$ or $v$ as in Eq.~(\ref{E:ENewt}), however the
recognition that we can set the relative velocity $v = 0$ provides the key for calculating the relativistic
distribution $d^3m/dEdL_zdQ$.

It is often convenient to use Fermi normal coordinates when calculating quantities near a point (such as the
center of mass of a tidally disrupting star) that moves along a Kerr geodesic.  We review the use of these
coordinates in Appendix~\ref{A:Fermi}.  If a fluid element is located at a fixed position $X^i$ in Fermi
normal coordinates, its relative velocity with respect to the origin (the center of mass) will vanish
\cite{Wald:1984rg}:
\begin{equation}
v^\alpha = \lambda_{0}^\beta \nabla_\beta (X^i \lambda_{i}^\alpha) = 
X^i \lambda_{0}^\beta \nabla_\beta \lambda_{i}^\alpha = 0~.
\end{equation}
The final equality follows because our basis vectors $\lambda_{A}^\alpha$ were constructed to be parallel
transported along the central geodesic ($\lambda_{0}^\beta \nabla_\beta \lambda_{i}^\alpha = 0$).  The
difference in specific energy between this fluid element and one located at the origin is given by
\begin{eqnarray} \label{E:delErel}
\Delta E &=&  X^i \lambda_{i}^\alpha \nabla_\alpha E \nonumber \\
&=& X^i \lambda_{i}^\alpha \nabla_\alpha \left[ -g_{\beta\gamma} \lambda_{0}^\beta
\left( \frac{\partial}{\partial t} \right)^\gamma \right] \nonumber \\
&=& -g_{\beta\gamma} \lambda_{0}^\beta X^i \lambda_{i}^\alpha \nabla_\alpha
\left( \frac{\partial}{\partial t} \right)^\gamma \nonumber \\
&=& -g_{\beta\gamma} \lambda_{0}^\beta X^i \lambda_{i}^\alpha \Gamma_{\alpha t}^\gamma~.
\end{eqnarray}
The second equality follows from the definition of $E$ in Eq.~(\ref{E:Edef}).  The third equality follows from
the compatibility of the metric with the derivative operator, the commutation of coordinate vector fields
($\lambda_{i}^\alpha \nabla_\alpha \lambda_{0}^\beta = \lambda_{0}^\alpha \nabla_\alpha
\lambda_{i}^\beta$) \cite{Wald:1984rg}, and the parallel transport of the basis vectors.  The final equality
follows from the definition of the Christoffel symbols.  Similar calculations show that
\begin{eqnarray} \label{E:delLrel}
\Delta L_z &=&  X^i \lambda_{i}^\alpha \nabla_\alpha L_z \nonumber \\
&=& g_{\beta\gamma} \lambda_{0}^\beta X^i \lambda_{i}^\alpha \Gamma_{\alpha\phi}^\gamma~,
\end{eqnarray}
and
\begin{eqnarray} \label{E:delKrel}
\Delta K &=&  X^i \lambda_{i}^\alpha \nabla_\alpha K \nonumber \\
&=& 2X^i[\lambda_{0}^\alpha \lambda_{0}^\beta \lambda_{i}^\gamma \nabla_\gamma
(\Sigma l_\alpha n_\beta) - r\lambda_{i}^r]~,
\end{eqnarray}
where $K$ is defined in Eq.~(\ref{E:Qdef}) and $l^\alpha$ and $n^\alpha$ are given by Eq.~(\ref{E:nullV}). 
We can then compute the difference
\begin{equation} \label{E:delQ}
\Delta Q = \Delta K - 2(L_z - aE)(\Delta L_z - a\Delta E)
\end{equation}
in the Carter constant $Q$ of the fluid element compared to the center of mass.

Our prescription for calculating the mass fallback rate in the relativistic limit is as follows:

\begin{itemize}

\item[(1)] Choose an orbit for the tidally disrupting star characterized by its 4-velocity $\lambda_{0}^\alpha$ at
pericenter and constants of motion $E$, $L_z$, and $Q$.

\item[(2)] Determine the density profile $\rho(r)$ of the star from its mass $m_\ast$, radius $R_\ast$, and
adiabatic index $\gamma$ by solving the Lane-Emden equation (\ref{E:LaEm}).

\item[(3)] Use $\rho(r)$ to determine a distribution of positions $X^i$ for the fluid elements in Fermi normal
coordinates.

\item[(4)] Use Eqs.~(\ref{E:delErel}), (\ref{E:delLrel}), (\ref{E:delKrel}), and (\ref{E:newx}) to determine the
positions $x^\alpha$ of the fluid elements in Boyer-Lindquist coordinates and their constants of motion
$E + \Delta E$, $L_z + \Delta L_z$, and $Q + \Delta Q$.

\item[(5)] Determine from their positions and constants of motion whether each fluid element is on a geodesic
that plunges directly into the event horizon, escapes to infinity, or becomes bound to the SBH.

\item[(6)] If a fluid element is on a bound orbit, integrate the equations of motion (\ref{E:EOM}) out to
apocenter beginning from $t =0$; the final time $t_f$ will be half of the radial period.

\item[(7)] Create a histogram of the resulting distribution of orbital periods to determine the fallback accretion
rate $dm/dt$, similar to that depicted in Fig.~\ref{F:dmdt_Lod} in the Newtonian limit.

\end{itemize}

\section{Results} \label{S:res}

We begin by presenting our predictions for the distribution $d^3m/dEdL_zdQ$ in Sec.~\ref{SS:dist}; this is
intermediate step (4) in the prescription given above.  While this provides some insight into
how SBH spin affects the tidal debris, those interested only in the corresponding fallback accretion rates
$dm/dt$ can skip ahead to Sec.~\ref{SS:fallback}.

\subsection{Distribution of orbital constants} \label{SS:dist}

Eqs.~(\ref{E:delErel}) - (\ref{E:delQ}) reveal that $\Delta E$, $\Delta L_z$, and $\Delta Q$ for each fluid
element are inner products of the position $X^i$ of that fluid element in Fermi normal coordinates with
covariant derivatives in the direction of the spatial basis vectors $\lambda_{i}^\alpha$.  In the Newtonian
limit, these quantities are inner products of $\mathbf{r} - \mathbf{r}_p$ with flat-space gradients.  We have
already seen this in Eq.~(\ref{E:Ecorr}), the Newtonian limit of Eq.~(\ref{E:delErel}).  This implies that
the projected distributions $dm/dE$, $dm/dL_z$, and $dm/dQ$ will have the {\it exact same shape} as the
Newtonian distribution shown in Fig.~\ref{F:dmdE_Lod}.  The width of the relativistic distribution $dm/dE$
is defined as 
\begin{equation} \label{E:sigE}
\sigma_E \equiv |\lambda_{i}^\alpha \nabla_\alpha E| R_\ast~;
\end{equation}
the widths $\sigma_{L_z}$, and $\sigma_Q$ are found by replacing $E$ with $L_z$ or $Q$ in
Eq.~(\ref{E:sigE}).  These widths depend on the SBH mass $M$, dimensionless spin $a/M$, and the orbit of
the tidally disrupted star.  We explore this dependence in this section.

\begin{figure}[t!]
\begin{center}
\includegraphics[width=3.5in]{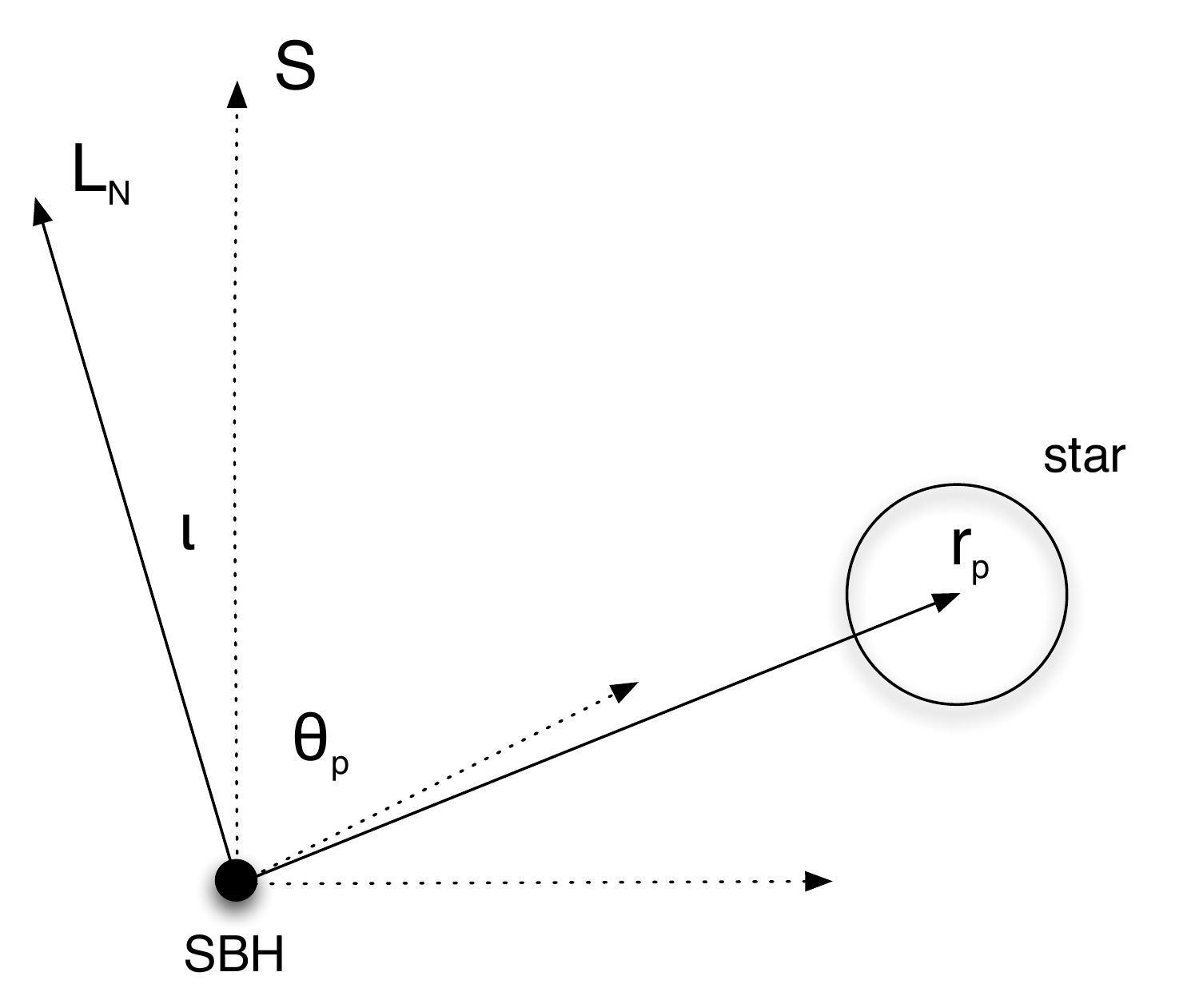}
\end{center}
\caption{The SBH is located at the origin with its spin $\mathbf{S}$ pointing along the z axis.  The inclination
$\iota$ is the angle between $\mathbf{S}$ and the star's orbital angular momentum $\mathbf{L}_N$, and
$\theta_p$ is the angle between $\mathbf{S}$ and the position $\mathbf{r}_p$ of the star at pericenter.}
\label{F:orbit}
\end{figure}

In the Newtonian limit, $\sigma_E \to E_{\rm tid}$ given by Eq.~(\ref{E:Etid}).  We did not calculate the
corresponding widths $L_{z,{\rm tid}}$ and $Q_{\rm tid}$ of $dm/dL_z$ and $dm/dQ$ in Sec.~\ref{S:Newt}, as
these quantities did not enter into the Newtonian orbital period given by Eq.~(\ref{E:binper}).  As the radial
period for bound orbits in the Kerr spacetime does depend on $L_z$ and $Q$, we now calculate these
widths.  In the Newtonian limit,
\begin{eqnarray} \label{E:DelLNewt}
\boldsymbol{\Delta}\mathbf{L}_N &=&  (\mathbf{r} - \mathbf{r}_p) \times \mathbf{v} \nonumber \\
&=& \left( \frac{2GM}{r_p} \right)^{1/2} \{ [(\mathbf{r} - \mathbf{r}_p) \cdot \mathbf{\hat{r}}_p] \mathbf{\hat{L}}_N
- [(\mathbf{r} - \mathbf{r}_p) \cdot \mathbf{\hat{L}}_N] \mathbf{\hat{r}}_p \} \quad \quad
\end{eqnarray}
implying that
\begin{eqnarray} \label{E:DelLzNewt}
\Delta L_z &=& \boldsymbol{\Delta}\mathbf{L}_N \cdot \mathbf{\hat{z}} \nonumber \\
&=& \left( \frac{2GM}{r_p} \right)^{1/2} (\mathbf{\hat{r}}_p \cos \iota - \mathbf{\hat{L}}_N \cos \theta_p) \cdot
(\mathbf{r} - \mathbf{r}_p) \quad \quad
\end{eqnarray}
and
\begin{eqnarray} \label{E:DelQNewt}
\Delta Q &=& 2\boldsymbol{\Delta}\mathbf{L}_N \cdot [ (\mathbf{L}_N \cdot \mathbf{\hat{x}}) \mathbf{\hat{x}}
+ (\mathbf{L}_N \cdot \mathbf{\hat{y}}) \mathbf{\hat{y}} ] \nonumber \\
&=& 4GM (\mathbf{\hat{r}}_p \sin^2 \iota + \mathbf{\hat{L}}_N \cos \theta_p \cos \iota) \cdot
(\mathbf{r} - \mathbf{r}_p)~. \quad
\end{eqnarray}
where $\mathbf{L}_N$ is the orbital angular momentum and the angles $\theta_p$ and $\iota$ are defined
in Fig.~\ref{F:orbit}.   Eqs.~(\ref{E:DelLzNewt}) and (\ref{E:DelQNewt}) allow us to define the widths of the
$L_z$ and $Q$ distributions in the Newtonian limit:
\begin{subequations} \label{E:LQNewtwid}
\begin{eqnarray}
\label{E:LNewtwid}
L_{z, {\rm tid}} &=& \left( \frac{2GM}{r_p} \right)^{1/2} (\cos^2 \iota + \cos^2 \theta_p)^{1/2} R_\ast~, \\
\label{E:QNewtwid}
Q_{\rm tid} &=& 4GM (\sin^4 \iota + \cos^2 \theta_p \cos^2 \iota)^{1/2} R_\ast~. \quad
\end{eqnarray}
\end{subequations}
Examining the ratio $\sigma_X/X_{\rm tid}$ of the widths of the distribution $dm/dX$ of a quantity $X$ in the
relativistic and Newtonian limits will allow us to assess the accuracy of the Newtonian calculation.

\subsubsection{Equatorial orbits} \label{SSS:EqCon}

\begin{figure}[t!]
\begin{center}
\includegraphics[width=3.5in]{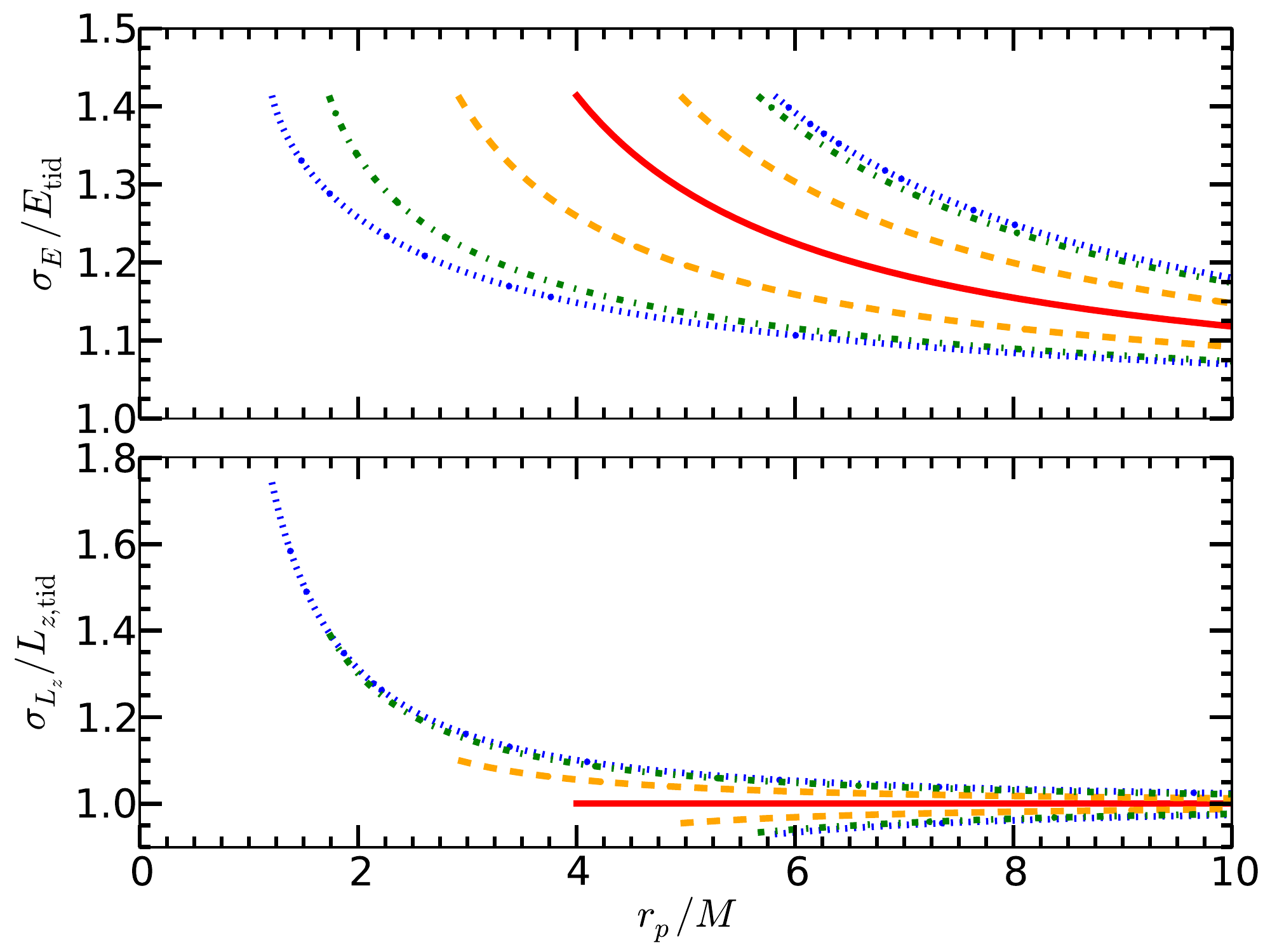}
\end{center}
\caption{The widths $\sigma_E$ and $\sigma_{L_z}$ of the projected distributions $dm/dE$ and $dm/dL_z$
as functions of the pericenter radius $r_p/M$.  These are normalized by the widths $E_{\rm tid}$ and
$L_{z, {\rm tid}}$ of the corresponding Newtonian distributions.  The solid red curves are for nonspinning
black holes, while the dashed orange, dot-dashed green, and dotted blue curves are for dimensionless spins
$a/M = 0.5$, 0.9, and 0.99 respectively.  For each value of the spin, the curve extending to smaller values of
$r_p$ corresponds to prograde ($\iota = 0^\circ$) orbits while the other curve corresponds to retrograde
($\iota = 180^\circ$) orbits.}
\label{F:EqCon}
\end{figure}

Equatorial orbits ($\theta_p = 90^\circ$) can be either prograde ($\iota = 0^\circ$) or retrograde
($\iota = 180^\circ$).  These orbits have $Q = 0$, and Eq.~(\ref{E:QNewtwid}) shows that the tidal debris will
also have $Q = 0$ in the Newtonian limit.  This holds for the relativistic calculation as well.  If we define
\begin{equation} \label{E:fr}
f_r \equiv 1 - \frac{2GM}{r_p}~,
\end{equation}
then for nonspinning SBHs,
\begin{subequations} \label{E:sigELNS}
\begin{eqnarray}
\label{E:sigENS}
\frac{\sigma_E}{E_{\rm tid}} &=& f_{r}^{-1/2}, \\
\label{E:sigLNS}
\frac{\sigma_{L_z}}{L_{z, {\rm tid}}} &=& \left[ \frac{\cos^2 \iota + f_{r}^{-1} \cos^2 \theta_p}{\cos^2
\iota + \cos^2 \theta_p} \right]^{1/2}.
\end{eqnarray}
\end{subequations}
We show these ratios for several additional values of the dimensionless spin $a/M$ in Fig.~\ref{F:EqCon}.
At fixed values of $r_p$, $\sigma_E$ decreases (increases) with $a/M$ on prograde (retrograde) orbits. 
However, at the radius $r_{\rm mb}$ of the marginally bound circular orbit (minimum value of $r_p$ for
each curve), $\sigma_E/E_{\rm tid} = \sqrt{2}$ independent of SBH spin.  By contrast, the ratio
$\sigma_{L_z}/L_{z, {\rm tid}}$ at $r_{\rm mb}$ depends sharply on $a/M$.  Although
$\sigma_{L_z}/L_{z, {\rm tid}} = 1$ all the way down to $r_{\rm mb} = 4M$ for nonspinning SBHs, it increases
to 2 (decreases to $\sim 0.929$) as $a/M \to 1$ for prograde (retrograde) orbits.

\subsubsection{Inclined orbits} \label{SSS:IncCon}

Inclined orbits have $0^\circ < \iota < 180^\circ$; for orbits with $L_z < 0$ it is convenient to define a
supplementary inclination
\begin{equation} \label{E:suppinc}
\cos \iota^\prime \equiv \frac{-L_z}{\sqrt{Q + L_{z}^2}}
\end{equation}
such that $\iota^\prime = 180^\circ - \iota$.  If we define a complementary angle $\theta_{p}^\prime \equiv
90^\circ - \theta_p$, Eq.~(\ref{E:dthetadt}) implies that for parabolic ($E = 1$) orbits
\begin{subequations} \label{E:thetarange}
\begin{eqnarray}
\label{E:prorange}
-\iota \leq &\theta_{p}^\prime& \leq +\iota \quad \quad~\iota \leq 90^\circ~, \\
\label{E:retrange}
-\iota^\prime \leq &\theta_{p}^\prime& \leq +\iota^\prime \quad \quad \iota > 90^\circ~.
\end{eqnarray}
\end{subequations}
The relativistic corrections $\sigma_X/X_{\rm tid}$ to the width of a distribution $dm/dX$ will depend on both
$\iota$ and $\theta_p$ for inclined orbits, in addition to $r_p/M$ and $a/M$.

\begin{figure}[t!]
\begin{center}
\includegraphics[width=3.5in]{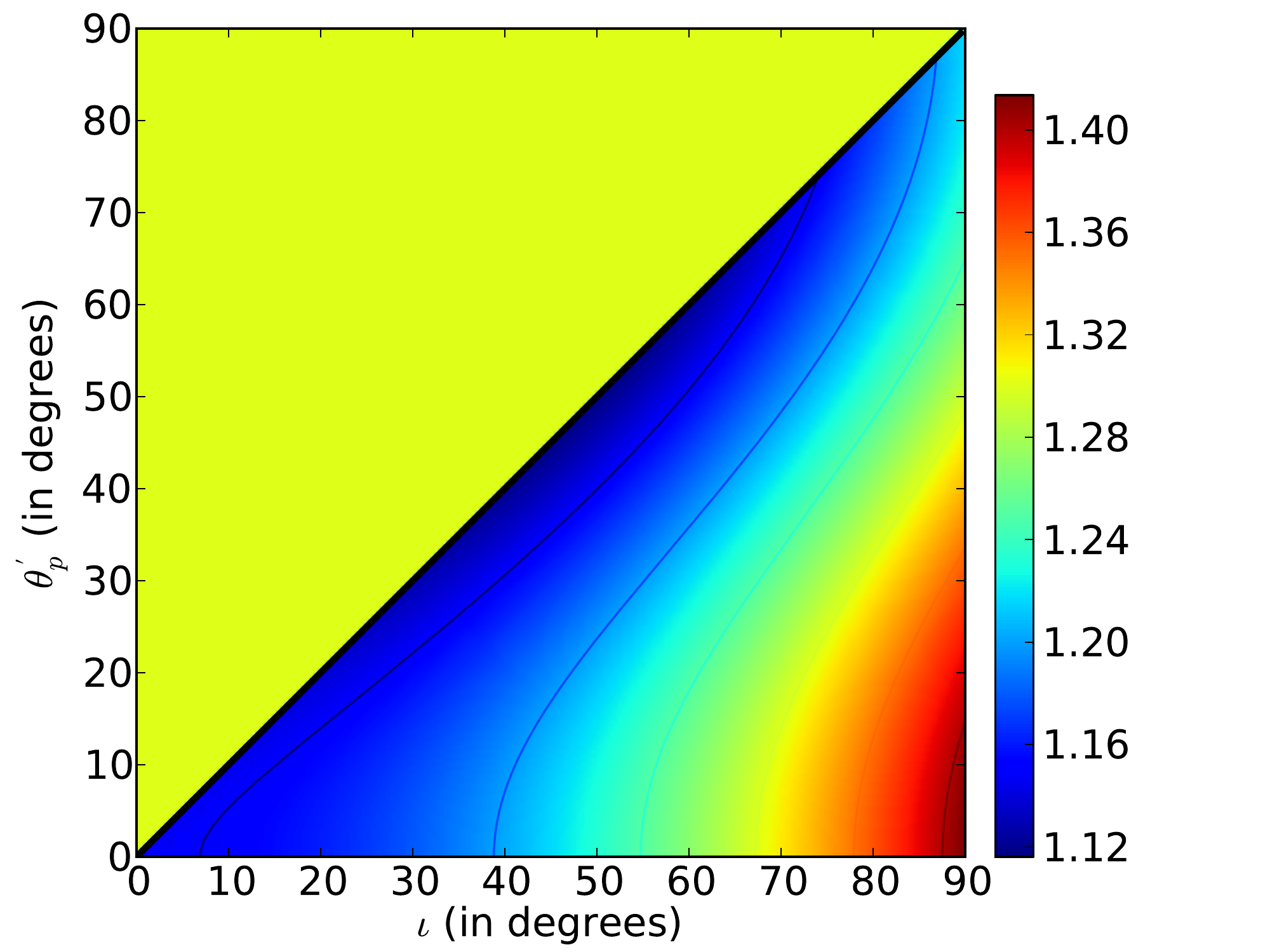}
\end{center}
\caption{The ratio $\sigma_E/E_{\rm tid}$ for orbits with $r_p = 4M$, $a/M = 0.99$, and $\iota \leq 90^\circ$
as a function of inclination $\iota$ and latitude $\theta_{p}^\prime = 90^\circ - \theta_p$.  The yellow region
above the diagonal is unphysical as it corresponds to a portion of parameter space forbidden by the
inequality
(\ref{E:prorange}).}
\label{F:ICE}
\end{figure}

We show $\sigma_E/E_{\rm tid}$ for inclined orbits with $r_p = 4M$, $a/M = 0.99$, and $\iota \leq 90^\circ$
in Fig.~\ref{F:ICE}.  The origin of this figure corresponds to the same orbit as that shown by the point at
$r_p = 4M$ on the lower dotted blue curve in the upper panel of Fig.~\ref{F:EqCon}.  We see that the
relativistic correction $\sigma_E/E_{\rm tid}$ is largest when tidal disruption occurs in the equatorial plane
($\theta_{p}^\prime = 0^\circ$) and increases with inclination $\iota$.  Although not apparent from this
figure, for higher spins and smaller radii the ratio $\sigma_E/E_{\rm tid}$ can exceed 1.5 on highly inclined
orbits, greater than the value $\sim \sqrt{2}$ found on the marginally bound orbits in Fig.~\ref{F:EqCon}.

\begin{figure}[t!]
\begin{center}
\includegraphics[width=3.5in]{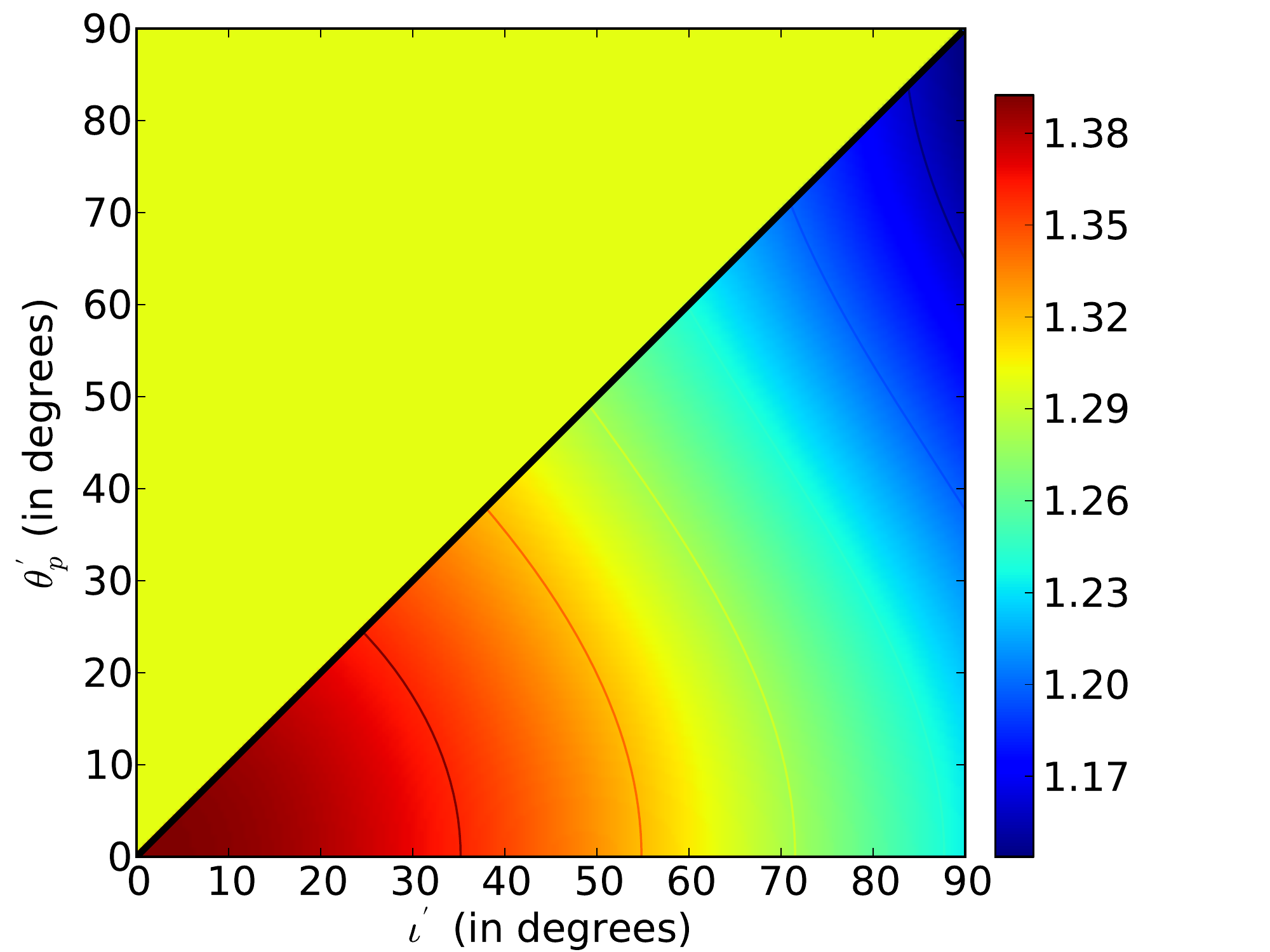}
\end{center}
\caption{The ratio $\sigma_E/E_{\rm tid}$ for orbits with $r_p = 6M$, $a/M = 0.99$, and $\iota > 90^\circ$ as
a function of supplementary inclination $\iota^\prime = 180^\circ - \iota$ and latitude $\theta_{p}^\prime$.
The portion of the plot above the diagonal is unphysical.}
\label{F:RCE}
\end{figure}

The ratio $\sigma_E/E_{\rm tid}$ for orbits with $r_p = 6M$, $a/M = 0.99$, and $\iota > 90^\circ$ is shown in
Fig.~\ref{F:RCE}.  We must use a larger value of $r_p$ than in Fig.~\ref{F:ICE}, as the marginally bound
orbits with $L_z < 0$ have larger radii.  The origin of this figure corresponds to the same orbit as that shown
by the point at $r_p = 6M$ on the upper dotted blue curve in the upper panel of Fig.~\ref{F:EqCon}.
As was the case for $\iota < 90^\circ$, the relativistic correction is greatest in the equatorial plane and
increases with inclination $\iota$.  Figs.~\ref{F:ICE} and \ref{F:RCE} are both consistent with our
impression from Fig.~\ref{F:EqCon} that the value of $\sigma_E/E_{\rm tid}$ largely depends on the
difference between $r_p$ and the radius of the marginally bound orbit for that spin and inclination.

\begin{figure}[t!]
\begin{center}
\includegraphics[width=3.5in]{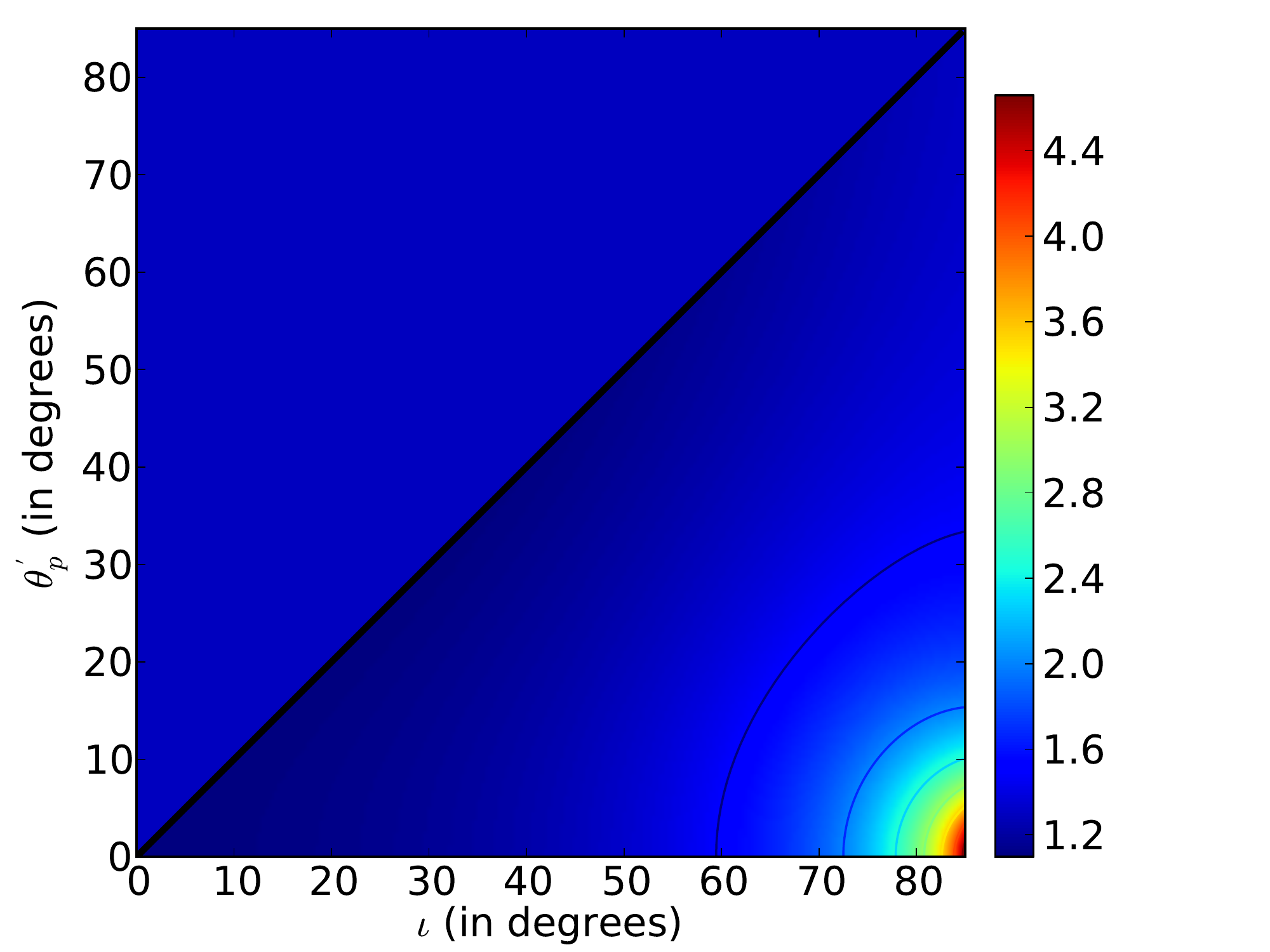}
\end{center}
\caption{The ratio $\sigma_{L_z}/L_{z, {\rm tid}}$ for orbits with $r_p = 4M$, $a/M = 0.99$, and $\iota \leq
85^\circ$ as a function of inclination $\iota$ and latitude $\theta_{p}^\prime$.  The region of the plot above the diagonal is unphysical.}
\label{F:ICL}
\end{figure}

\begin{figure}[t!]
\begin{center}
\includegraphics[width=3.5in]{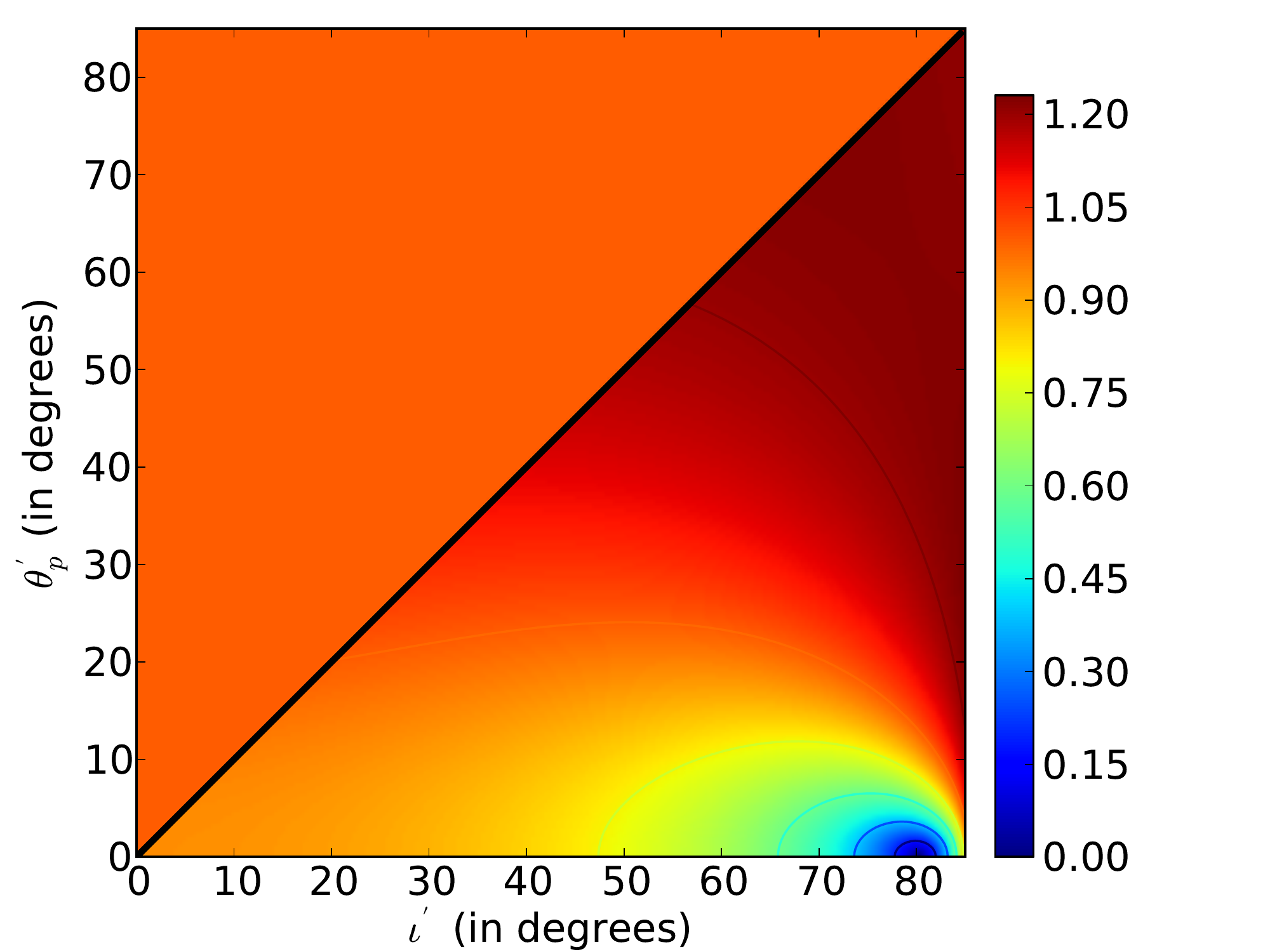}
\end{center}
\caption{The ratio $\sigma_{L_z}/L_{z, {\rm tid}}$ for orbits with $r_p = 6M$, $a/M = 0.99$, and $\iota \geq
95^\circ$ as a function of supplemental inclination $\iota^\prime = 180^\circ - \iota$ and latitude
$\theta_{p}^\prime$.  The region of the plot above the diagonal is unphysical.}
\label{F:RCL}
\end{figure}

While $E_{\rm tid}$ is independent of orbital inclination, both the numerator and denominator of the ratio
$\sigma_{L_z}/L_{z, {\rm tid}}$ depend on $\iota$ and $\theta_p$.  Eq.~(\ref{E:LNewtwid}) and
(\ref{E:sigLNS}) show that although both $\sigma_{L_z}$and $L_{z, {\rm tid}} \to 0$ as  $\iota \to 90^\circ$,
$\theta_p \to 90^\circ$ for $a/M = 0$, their ratio is undefined in this limit.  For $a/M > 0$, the location at which
$\sigma_{L_z} = 0$ shifts to $\iota_{0}^\prime(a/M) < 90^\circ$, $\theta_p = 90^\circ$ while the Newtonian
denominator $L_{z, {\rm tid}}$ of Eq.~(\ref{E:LNewtwid}) remains unchanged.  This implies that the ratio
$\sigma_{L_z}/L_{z, {\rm tid}}$ will have a pole at $\iota = 90^\circ$, $\theta_{p}^\prime = 0^\circ$, as can be
seen in Fig.~\ref{F:ICL} where we show this ratio for orbits with $r_p = 4M$, $a/M = 0.99$, and
$\iota \leq 85^\circ$.  The figure is cut off above $\iota = 85^\circ$ to avoid this pole.  The presence of this
pole is also apparent in the contour plot of $\sigma_{L_z}/L_{z, {\rm tid}}$ for orbits with $r_p = 6M$,
$a/M = 0.99$, and $\iota^\prime \leq 85^\circ$ shown in Fig.~\ref{F:RCL}.  We also see that
$\sigma_{L_z} \to 0$ as one approaches $\iota_0^\prime \simeq 80^\circ$, $\theta_{p}^\prime = 0^\circ$.

\begin{figure}[t!]
\begin{center}
\includegraphics[width=3.5in]{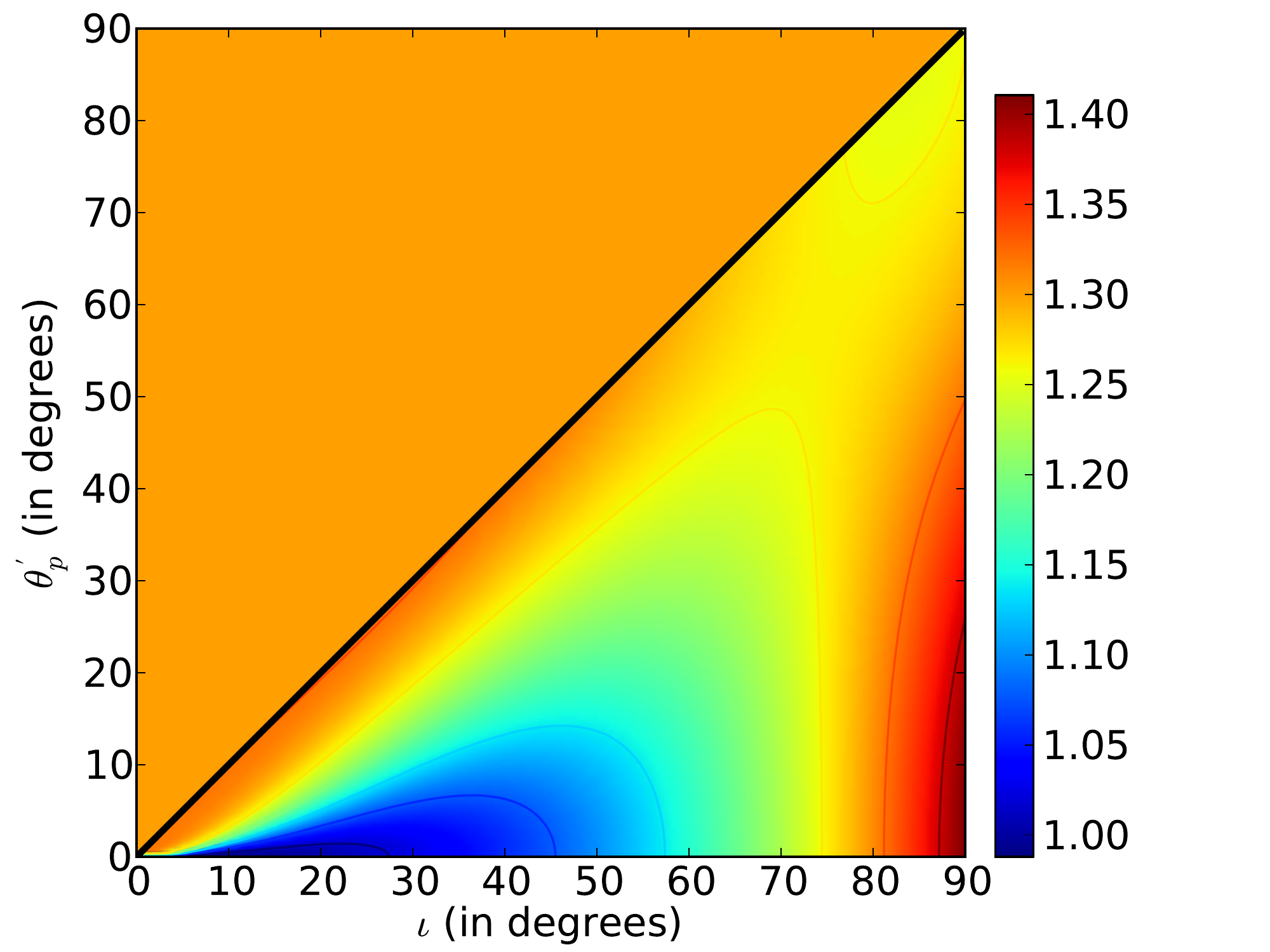}
\end{center}
\caption{The ratio $\sigma_Q/Q_{\rm tid}$ for orbits with $r_p = 4M$, $a/M = 0.99$, and $\iota \leq
90^\circ$ as a function of inclination $\iota$ and latitude $\theta_{p}^\prime$.  The region of the plot above the diagonal is unphysical.}
\label{F:ICQ}
\end{figure}

\begin{figure}[t!]
\begin{center}
\includegraphics[width=3.5in]{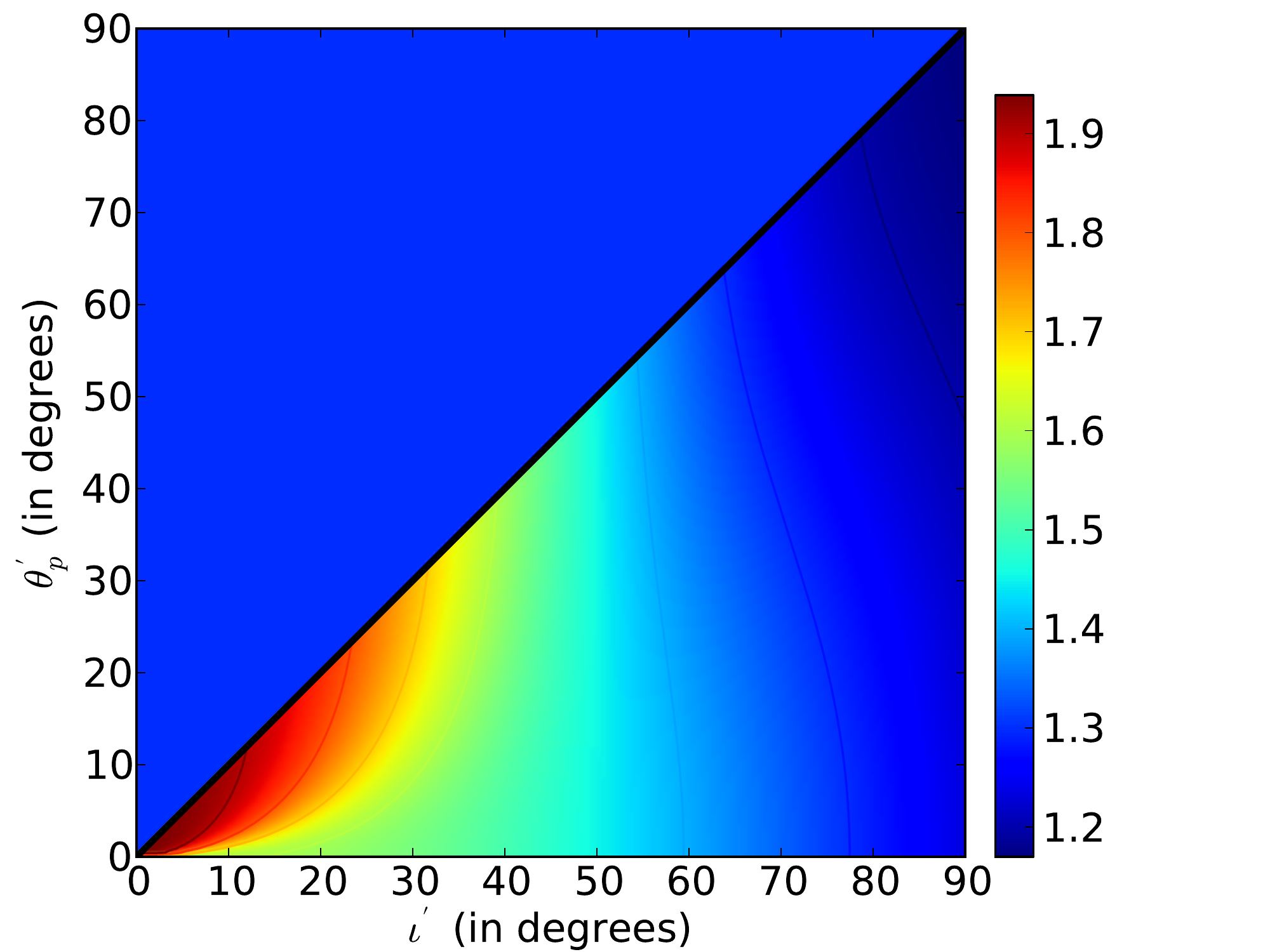}
\end{center}
\caption{The ratio $\sigma_Q/Q_{\rm tid}$ for orbits with $r_p = 6M$, $a/M = 0.99$, and $\iota \geq
90^\circ$ as a function of supplemental inclination $\iota^\prime = 180^\circ - \iota$ and latitude
$\theta_{p}^\prime$.  The region of the plot above the diagonal is unphysical.}
\label{F:RCQ}
\end{figure}

Finally we come to the width of distribution of the Carter constant $Q$, which for nonspinning SBHs is
given by
\begin{equation} \label{E:sigQNS}
\frac{\sigma_Q}{Q_{\rm tid}} = \left[ f_r^{-1} \frac{\sin^4 \iota + f_r^{-1} \cos^2 \theta_p \cos^2 \iota}
{\sin^4 \iota + \cos^2 \theta_p \cos^2 \iota} \right]^{1/2}.
\end{equation}
We show this ratio for orbits with $r_p = 4M$, $a/M = 0.99$, and $\iota \leq 90^\circ$ in Fig.~\ref{F:ICQ}.
This contour plot differs qualitatively from the nonspinning result (\ref{E:sigQNS}) in that
$\sigma_Q/Q_{\rm tid}$ is maximized at $\iota = 90^\circ$, $\theta_p^\prime = 0^\circ$ instead of near
$\iota = 0^\circ$, $\theta_p^\prime = 0^\circ$.  We show $\sigma_Q/Q_{\rm tid}$ for orbits with $r_p = 6M$,
$a/M = 0.99$, and $\iota \geq 90^\circ$ in Fig.~\ref{F:RCQ}.  This figure more closely resembles the
nonspinning result (\ref{E:sigQNS}), although $\sigma_Q^2$ no longer separates into two distinct terms, only
one of which is dependent on $\theta_p$.

\subsection{Fallback accretion rate} \label{SS:fallback}

We now present the fallback accretion rate $dm/dt$ for stellar tidal disruptions with different values of the
SBH mass $M$, SBH spin $a/M$, pericenter distance $r_p$, orbital inclination $\iota$, and polar angle
$\theta_p$ at pericenter.  We fix $m_\ast = M_\odot$, $R_\ast = R_\odot$, and $\gamma = 5/3$ throughout
this section, as our focus is on relativistic effects that are independent of these parameters.  We measure time
in units of $t_{\rm tid}$ given by Eq.~(\ref{E:delay}), and $r_p$ in units of $M$ (where $G = c = 1$).

\begin{figure}[t!]
\begin{center}
\includegraphics[width=3.5in]{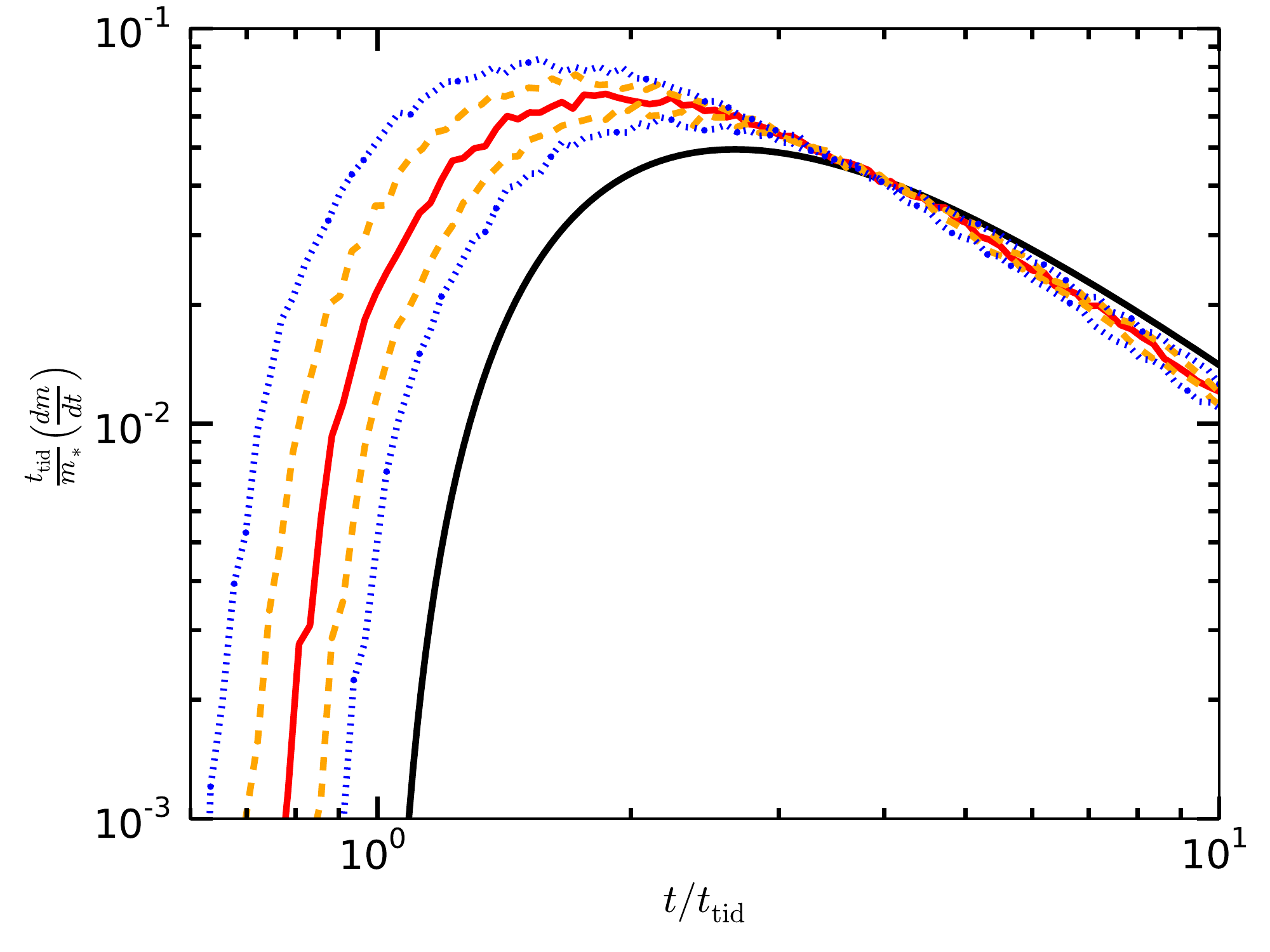}
\end{center}
\caption{The dimensionless fallback accretion rate $(t_{\rm tid}/m_\ast) dm/dt$ as a function of time
$t/t_{\rm tid}$ following the tidal disruption of a stars on equatorial orbits with $r_p = 6M$ by SBHs with
$M = 10^8 M_\odot$.  The smooth, solid black curve is the Newtonian prediction shown in
Fig.~\ref{F:dmdt_Lod}.  The solid red curve is the relativistic prediction for a non-spinning SBH.  The dashed
orange (dotted blue) curves correspond to spin magnitudes $a/M = 0.5~(0.99)$ respectively.  The curves of
each type with lower (higher) peak accretion rates correspond to orbital inclinations $\iota =
0^\circ~(180^\circ)$.} \label{F:r6M8E}
\end{figure}

\subsubsection{Equatorial orbits} \label{SSS:EqRates}

In Fig.~\ref{F:r6M8E}, we show the fallback accretion rate $dm/dt$ for TDEs of stars on equatorial
($\theta_p = 90^\circ$), parabolic ($E = 0$) orbits with $r_p = 6M$ by SBHs with $M = 10^8 M_\odot$.
The different curves correspond to
different values of the SBH spin and either prograde or retrograde orbits as described in the caption.  We see
that relativistic corrections can substantially affect the accretion rate for stars with such small pericenters.
The peak accretion rate $dm_{\rm peak}/dt$ can be almost twice as large as the Newtonian prediction, and
can occur at a time $t_{\rm peak}$ only half that of the Newtonian prediction.  SBH spin magnitude and
direction also strongly affect the accretion rate; large prograde ($\iota = 0^\circ$) spins reduce
$dm_{\rm peak}/dt$ and delay $t_{\rm peak}$, while large retrograde spins ($\iota = 180^\circ$) have the
opposite effect.  The jaggedness in the curves results from the finite number of particles ($N = 10^6$) and
finite proper time step ($\Delta \tau = 10^3 M$).  Each curve took approximately 15 minutes to prepare on a
single-processor 2.2~GHz laptop; better optimization and additional computational resources could allow a
more thorough exploration of parameter space with greater accuracy.

\begin{figure}[t!]
\begin{center}
\includegraphics[width=3.5in]{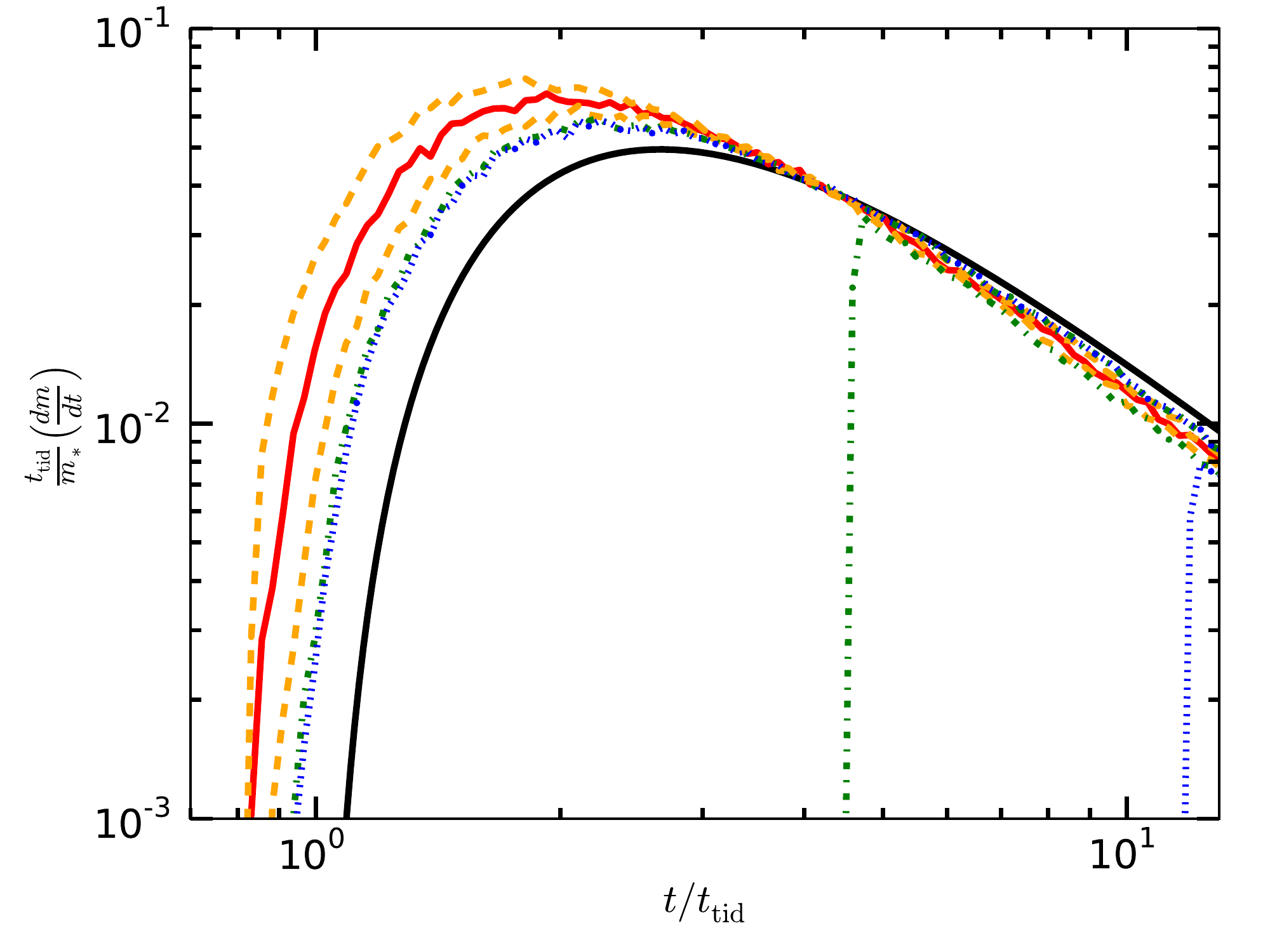}
\end{center}
\caption{The dimensionless fallback accretion rate $(t_{\rm tid}/m_\ast) dm/dt$ as a function of time
$t/t_{\rm tid}$ following the tidal disruption of a stars on equatorial orbits with $r_p = 6M$ by SBHs with
$M = 10^6 M_\odot$.  The smooth, solid black curve is the Newtonian prediction.  The solid red curve is the
relativistic prediction for a non-spinning SBH.  The dashed orange, dot-dashed green, and dotted blue
curves correspond to spin magnitudes $a/M$ of 0.5, 0.9, and 0.99 respectively.  The curves of each type with
higher (lower) accretion rates at late times correspond to orbital inclinations $\iota = 0^\circ~(180^\circ)$.} \label{F:r6M6E}
\end{figure}

In Fig.~\ref{F:r6M6E}, we again show the fallback accretion rate for equatorial orbits with $r_p = 6M$, but
have reduced the SBH mass from $10^8 M_\odot$ to $10^6 M_\odot$.  By measuring time in units of
$t_{\rm tid}$, we have eliminated the the dependence of the Newtonian prediction on $M$ and $r_p$.  The
magnitude of the relativistic effects is similar to that in Fig.~\ref{F:r6M8E} for the prograde orbits, but the
accretion rate has been sharply cut off at early times for the retrograde orbits.  This cutoff occurs because the
mostly tightly bound particles (those which would have the shortest orbital periods) have lost so much energy
$E$ and angular momentum $L_z$ that they are directly swallowed by the SBH event horizon.  This effect is
very small for $a/M = 0.5$, where only 0.021\% of the star's initial mass is directly swallowed (a slight
reduction in the retrograde dashed orange curve at the earliest times is just barely noticeable).  The cutoff is
much more pronounced for $a/M = 0.9$ and 0.99 shown by the dot-dashed green and dotted blue curves
respectively.  In these cases 21.4\% and 34.2\% of the initial stellar mass is directly captured.  This directly
captured material will have little opportunity to radiate, and should therefore not contribute to
observed TDE light curves.

In our dimensionless units, the Eddington accretion rate is
\begin{eqnarray} \label{E:dimEdd}
\frac{t_{\rm tid}}{m_\ast} \frac{dm_{\rm Edd}}{dt} &=& \frac{4\pi^2 m_p}{\eta m_\ast^2 \sigma_T c}
\beta^{-3} \left( \frac{GM^3 R_\ast^3}{2} \right)^{1/2} \nonumber \\
&=& 51.4 \left( \frac{\eta}{0.1} \right)^{-1} \left( \frac{m_\ast}{M_\odot} \right)^{-1}
\left( \frac{R_\ast}{R_\odot} \right)^{-3/2} \nonumber \\
&& \times \left( \frac{r_p}{6M} \right)^3 \left( \frac{M}{10^8 M_\odot} \right)^{7/2}
\end{eqnarray}
where $m_p$ is the proton mass, $\eta$ is the radiative efficiency, and $\sigma_T$ is the Thomson cross
section.  While the peak accretion rates for $10^8 M_\odot$ SBHs as shown in Fig.~\ref{F:r6M8E} are
comfortably sub-Eddington, those for $10^6 M_\odot$ SBHs as shown in Fig.~\ref{F:r6M6E} are highly
super-Eddington.  It is therefore unlikely that the observed TDE luminosity for $M = 10^6 M_\odot$
will track the fallback accretion rate at early times which are most sensitive to SBH spin.  More massive SBHs
have peak accretion rates below the Eddington limit, but are only capable of tidally disrupting stars when
\begin{equation} \label{E:rtid}
\frac{r_p}{M} \lesssim \frac{r_{\rm tid}}{M} \simeq 47.1 \left( \frac{M}{10^6 M_\odot} \right)^{-2/3}
\left( \frac{m_\ast}{M_\odot} \right)^{-1/3} \left( \frac{R_\ast}{R_\odot} \right)~.
\end{equation}
SBHs as massive as $10^8 M_\odot$ will directly capture many stars without tidally disrupting them,
although relativistic effects allow TDEs for $M \lesssim 10^9 M_\odot$ despite this Newtonian estimate
\cite{Beloborodov:1992,Ivanov:2005se,Kesden:2011ee}.  These considerations
suggest that $M \simeq 10^7 M_\odot$ may be a sweet spot for attempts to constrain SBH spins with TDE
light curves, which is fortuitous since SBHs of this mass are fairly ubiquitous at the modest redshifts where
TDEs are readily observable.

\begin{figure}[t!]
\begin{center}
\includegraphics[width=3.5in]{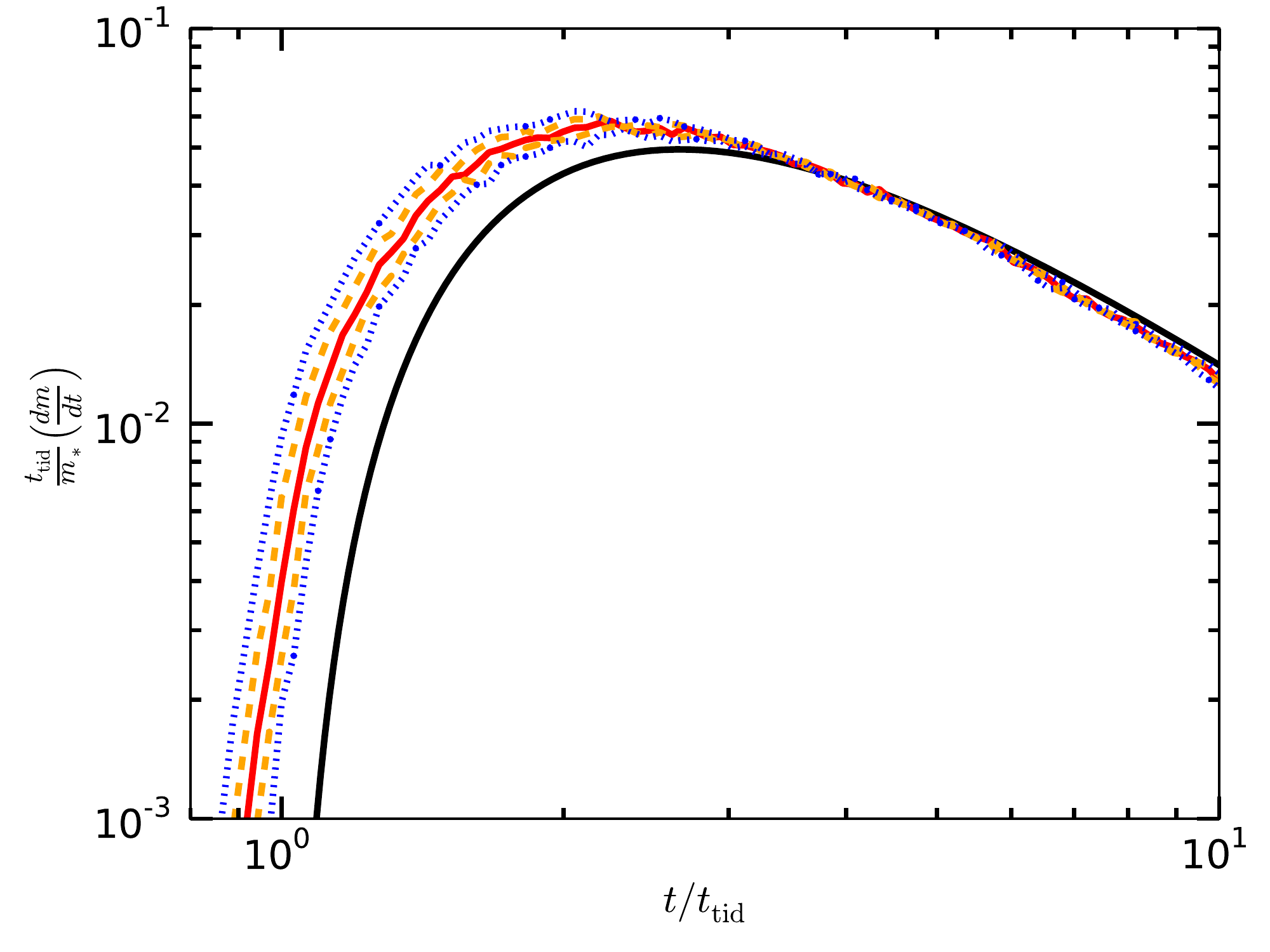}
\end{center}
\caption{The dimensionless fallback accretion rate $(t_{\rm tid}/m_\ast) dm/dt$ as a function of time
$t/t_{\rm tid}$ following the tidal disruption of a stars on equatorial orbits with $r_p = 12M$ by SBHs with
$M = 10^7 M_\odot$.  The smooth, solid black curve is the Newtonian prediction.  The solid red curve is the
relativistic prediction for a non-spinning SBH.  The dashed orange (dotted blue) curves correspond to spin
magnitudes $a/M$ of 0.5~(0.99).  The curves of each type with lower (higher) peak accretion rates
correspond to orbital inclinations $\iota = 0^\circ~(180^\circ)$.} \label{F:r12M7E}
\end{figure}

We show the fallback accretion rates for TDEs by a SBH with $M = 10^7 M_\odot$ in Fig.~\ref{F:r12M7E}.
The pericenter for the TDEs shown in this figure has been increased to $r_p = 12 M$ to show how relativistic
effects fall off with distance.  Comparing this figure to Fig.~\ref{F:r6M8E}, we see that relativistic corrections
increase $dm_{\rm peak}/dt$ and reduce $t_{\rm peak}$ by at most $\sim 25\%$, and SBH spin induces
similar corrections to these quantities at the $\sim 25\%$ level.

\subsubsection{Inclined orbits} \label{SSS:IncRates}

\begin{figure}[t!]
\begin{center}
\includegraphics[width=3.5in]{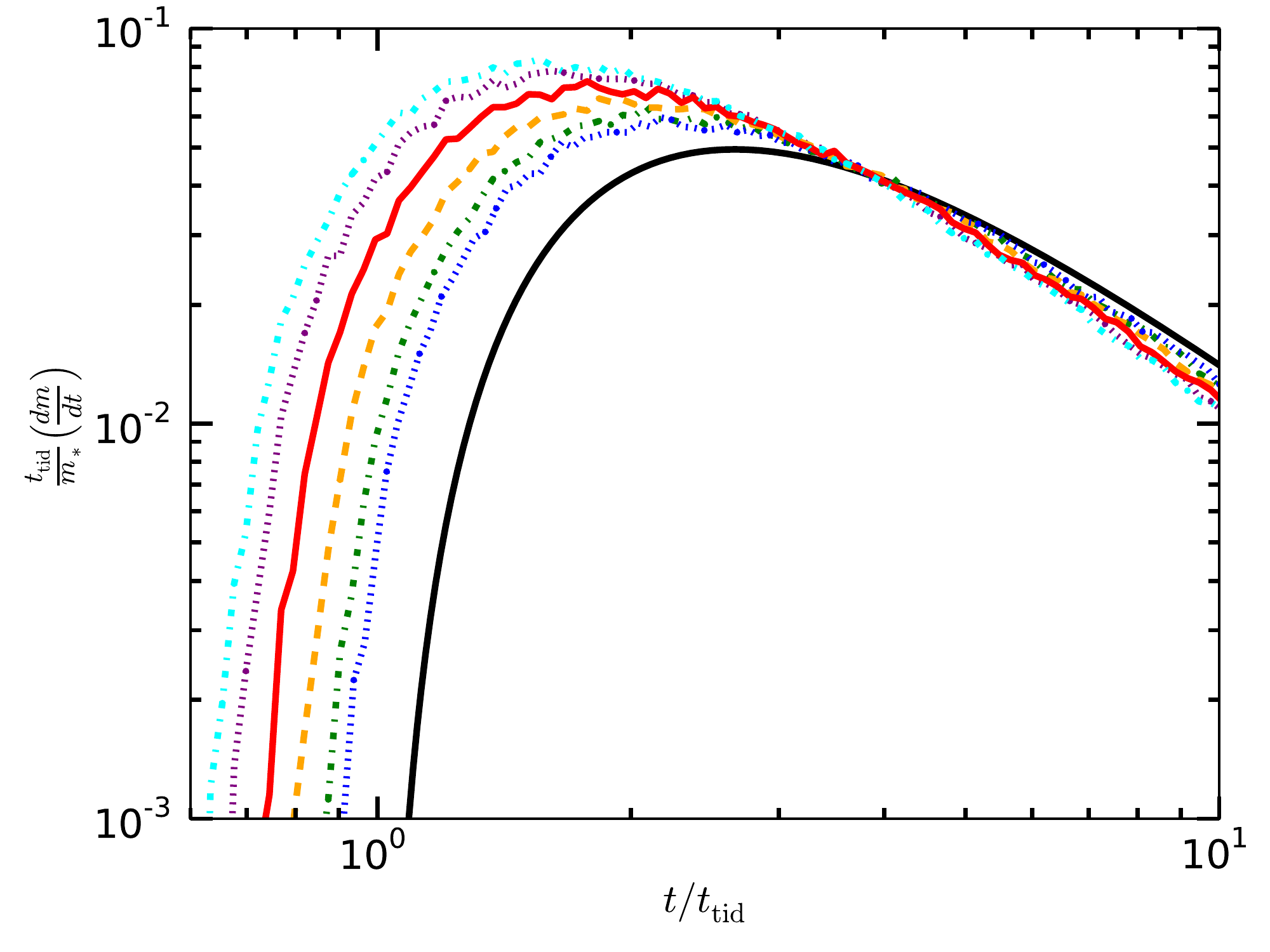}
\end{center}
\caption{The dimensionless fallback accretion rate $(t_{\rm tid}/m_\ast)dm/dt$ as a function of time
$t/t_{\rm tid}$ following the tidal disruption of a stars on inclined orbits with $r_p = 6M$ by SBHs with
$M = 10^8 M_\odot$ and $a/M = 0.99$.  The smooth, solid black curve is the Newtonian prediction.  The
remaining curves correspond to relativistic predictions with polar angle $\theta_p = 90^\circ$ but different
values of the orbital inclination $\iota$.  The peak accretion rate $dm_{\rm peak}/dt$ increases with $\iota$;
with increasing values of $dm_{\rm peak}/dt$, the dotted blue, dot-dashed green, dashed orange, solid red,
dotted purple, and dot-dashed cyan curves correspond to $\iota = 0^\circ$, $45^\circ$, $75^\circ$,
$105^\circ$, $135^\circ$, and $180^\circ$.} \label{F:r6M8I}
\end{figure}

Tidally disrupted stars are scattered onto their initial orbits at large radii where the influence of the SBH spin
is negligible.  One should therefore expect a flat distribution in $\cos \iota$ in the range $-1 \leq \cos \iota
\leq 1$.  We now consider how the fallback accretion rate depends on $\iota$ and the polar angle $\theta_p$
at pericenter for inclined orbits.  We show $dm/dt$ for fixed spin magnitude $a/M = 0.99$ and
$\theta_p = 90^\circ$ but different orbital inclinations $\iota$ in Fig.~\ref{F:r6M8I}.  The Newtonian
prediction in this figure is identical to that in Fig.~\ref{F:r6M8E}, and the dotted blue (dot-dashed cyan) curves
corresponding to $\iota = 0^\circ~(180^\circ)$ are identical to the dotted blue curves in Fig.~\ref{F:r6M8E}.
The remaining curves correspond to inclined orbits that were discussed in detail in Sec.~\ref{SSS:IncCon}.
In terms of the relativistic correction $\sigma_E/E_{\rm tid}$ to the width of the energy distribution, increasing
$\iota$ from $0^\circ$ to $180^\circ$ while keeping $\theta_p = 90^\circ$ fixed corresponds to moving from
left to right along the x-axis of Fig.~\ref{F:ICE}, then moving from right to left along the x-axis of
Fig.~\ref{F:RCE}.  The ratio $\sigma_E/E_{\rm tid}$ increases with $\iota$ along this path, qualitatively
implying that tidal debris falls deeper into the SBH's potential well with shorter radial periods $t/M$.  Although
in the relativistic limit the radial period depends on $L_z$ and $Q$ in addition to $E$, consideration of $E$
alone seems to illustrate the qualitative dependence on $\iota$.

\begin{figure}[t!]
\begin{center}
\includegraphics[width=3.5in]{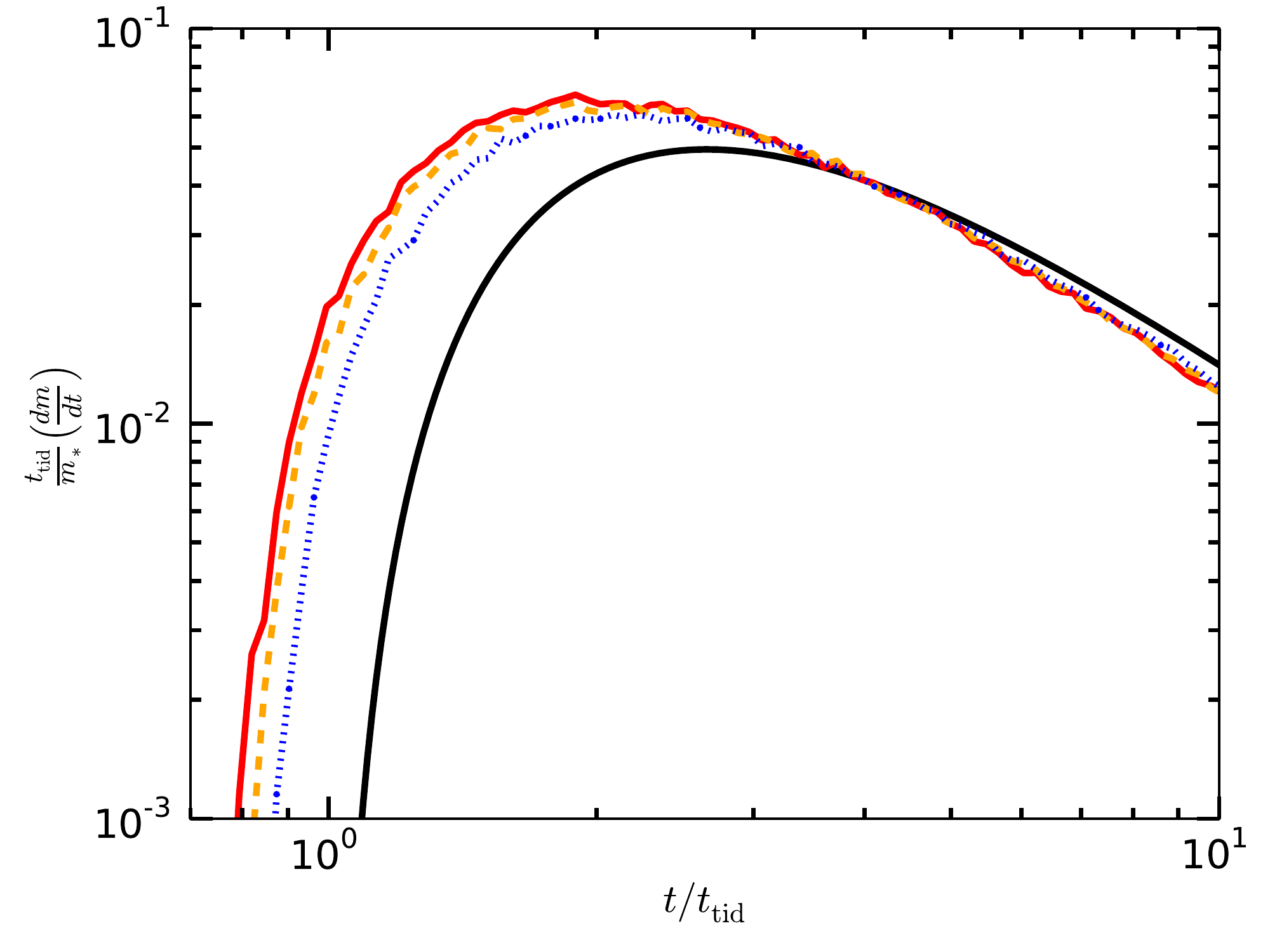}
\end{center}
\caption{The dimensionless fallback accretion rate $(t_{\rm tid}/m_\ast) dm/dt$ as a function of time
$t/t_{\rm tid}$ following the tidal disruption of a stars on inclined orbits with $r_p = 6M$ by SBHs with
$M = 10^8 M_\odot$ and $a/M = 0.99$.  The smooth, solid black curve is the Newtonian prediction.  The
remaining curves correspond to relativistic predictions with inclination $\iota = 80^\circ$ but different
values of the polar angle at pericenter $\theta_p$.  The peak accretion rate $dm_{\rm peak}/dt$ increases
with $\theta_p$; with increasing values of $dm_{\rm peak}/dt$, the dotted blue, dashed orange, and solid red,
curves correspond to $\theta_p = 30^\circ$, $60^\circ$, and $90^\circ$.} \label{F:r6M8P}
\end{figure}

We show the dependence of the fallback accretion rate on the polar angle at pericenter $\theta_p$ in
Fig.~\ref{F:r6M8P}.  As in the previous figure, Fig.~\ref{F:r6M8P} depicts TDEs of stars with pericenter
$r_p = 6M$ by a SBH with $M = 10^8 M_\odot$ and $a/M = 0.99$.  The orbital inclination is fixed at
$\iota = 80^\circ$, but the latitude $\theta^{\prime}_p = 90^\circ - \theta_p$ is varied from $0^\circ$ to
$60^\circ$.  In terms of the ratio $\sigma_E/E_{\rm tid}$, this corresponds to moving upwards along the
vertical line $\iota = 80^\circ$ in Fig.~\ref{F:ICE}.  The ratio $\sigma_E/E_{\rm tid}$ decreases as we move
along this path, implying that the tidal debris becomes less tightly bound to the SBH with longer radial
periods $t/M$.  This qualitatively accounts for the trend seen in Fig.~\ref{F:r6M8P}, that $dm/dt$ is shifted to
later times as $\theta_p$ decreases ($\theta^{\prime}_p$ increases).  Although not shown, this same trend
holds for $\iota > 90^\circ$ as one would expect from consideration of Fig.~\ref{F:RCE}.  Our results for the
peak accretion rate $dm_{\rm peak}/dt$ and the time $t_{\rm peak}$ at which this peak occurs are
summarized in Table~\ref{T:res}.

\begin{center}
\begin{table}[t!]
\begin{tabular}{ccccccc}
\hline
\hline
$M$ &$a/M$ &$r_p$ &$\iota$ &$\theta_p$ &$\frac{t_{\rm peak}}{t_{\rm tid}}$
&$\frac{t_{\rm tid}}{m_\ast} \frac{dm_{\rm peak}}{dt}$\\ 
\hline
\hline
$10^8 M_\odot$   &0.0    &$6M$    &$0^\circ$    &$90^\circ$    &1.87    &0.0683   \\
$10^8 M_\odot$   &0.5    &$6M$    &$0^\circ$    &$90^\circ$    &2.04    &0.0645   \\
$10^8 M_\odot$   &0.5    &$6M$    &$180^\circ$    &$90^\circ$    &1.72    &0.0766   \\
$10^8 M_\odot$   &0.99    &$6M$    &$0^\circ$    &$90^\circ$    &2.17    &0.0596   \\
$10^8 M_\odot$   &0.99    &$6M$    &$180^\circ$    &$90^\circ$    &1.56    &0.0838   \\
$10^6 M_\odot$   &0.0    &$6M$    &$0^\circ$    &$90^\circ$    &1.92    &0.0685   \\
$10^6 M_\odot$   &0.5    &$6M$    &$0^\circ$    &$90^\circ$    &2.11    &0.0638   \\
$10^6 M_\odot$   &0.5    &$6M$    &$180^\circ$    &$90^\circ$    &1.81    &0.0747   \\
$10^6 M_\odot$   &0.9    &$6M$    &$0^\circ$    &$90^\circ$    &2.24    &0.0595   \\
$10^6 M_\odot$   &0.9    &$6M$    &$180^\circ$    &$90^\circ$    &4.73    &0.0335   \\
$10^6 M_\odot$   &0.99    &$6M$    &$0^\circ$    &$90^\circ$    &2.24    &0.0585   \\
$10^6 M_\odot$   &0.99    &$6M$    &$180^\circ$    &$90^\circ$    &12.33    &0.0079   \\
$10^7 M_\odot$   &0.0    &$12M$    &$0^\circ$    &$90^\circ$    &2.25    &0.0583   \\
$10^7 M_\odot$   &0.5    &$12M$    &$0^\circ$    &$90^\circ$    &2.28    &0.0564   \\
$10^7 M_\odot$   &0.5    &$12M$    &$180^\circ$    &$90^\circ$    &2.17    &0.0599   \\
$10^7 M_\odot$   &0.99    &$12M$    &$0^\circ$    &$90^\circ$    &2.31    &0.0558   \\
$10^7 M_\odot$   &0.99    &$12M$    &$180^\circ$    &$90^\circ$    &2.05    &0.0617   \\
$10^8 M_\odot$   &0.99    &$6M$    &$45^\circ$    &$90^\circ$    &2.09    &0.0636   \\
$10^8 M_\odot$   &0.99    &$6M$    &$75^\circ$    &$90^\circ$    &1.89    &0.0665   \\
$10^8 M_\odot$   &0.99    &$6M$    &$105^\circ$    &$90^\circ$    &1.78    &0.0735   \\
$10^8 M_\odot$   &0.99    &$6M$    &$135^\circ$    &$90^\circ$    &1.61    &0.0781   \\
$10^8 M_\odot$   &0.99    &$6M$    &$80^\circ$    &$30^\circ$    &2.15    &0.0606   \\
$10^8 M_\odot$   &0.99    &$6M$    &$80^\circ$    &$60^\circ$    &2.02    &0.0651   \\
$10^8 M_\odot$   &0.99    &$6M$    &$80^\circ$    &$90^\circ$    &1.89    &0.0680   \\
\hline
\hline
\end{tabular}
\caption{Peak accretion rate $dm_{\rm peak}/dt$ and the time $t_{\rm peak}$ at which this peak occurs for
different values of the SBH mass $M$, spin $a/M$, pericenter $r_p$, orbital inclination $\iota$, and polar
angle $\theta_p$ at pericenter.}
\label{T:res}  
\end{table}
\end{center}

\section{Discussion} \label{S:disc}

Our primary goals in this paper have been to establish a theoretical framework for investigating stellar tidal
disruption deep in the relativistic regime and to conduct a preliminary survey of the many dimensional
parameter space associated with this problem.  Our starting point for this framework was the Newtonian
model proposed in LKP09, in which a star's orbit and density profile were used to determine the distribution
of orbital energy of the tidal debris and hence the fallback accretion rate.  The gravitational potential of a
Newtonian point mass is spherically symmetric, implying that for a given stellar profile the only physically
significant parameters are the SBH mass $M$ and the pericenter $r_p$.  The Kerr spacetime however is
only axisymmetric, increasing the number of required orbital parameters to include the SBH spin $a/M$, the
orbital inclination $\iota$, and the polar angle at pericenter $\theta_p$.  The radial period of the
tidal debris now depends on the angular momentum $L_z$ and Carter constant $Q$ in addition to the
energy $E$.  Fermi normal coordinates allow us to determine the distribution of these orbital constants for a
given stellar profile and orbit, and the Killing vectors and tensor of the Kerr spacetime provide a series of
first-order differential equations for the orbital motion of the tidal debris given the orbital constants.
Combining these ingredients according to the prescription given at the end of Sec.~\ref{S:rel}, we can
calculate the fallback accretion rate $dm/dt$ as a function of time $t$.

If $dm/dt$ is greater than the Eddington accretion rate, as will be the case for $M \lesssim 10^7 M_\odot$
and $t \simeq t_{\rm peak}$, the emission may be dominated by powerful super-Eddington outflows whose
luminosity will not be proportional to $dm/dt$ \cite{Strubbe:2009qs}.  At later times and for larger SBH
masses, the luminosity will be dominated by an accretion disk about the SBH, whose emission may be
partially reprocessed by unbound tidal debris.  The bolometric luminosity of this accretion disk should be
proportional to $dm/dt$, although the finite extent of the disk implies that the emission spans a fairly narrow
range of temperatures \cite{Lodato:2010xs}.  Monochromatic light curves may therefore not be proportional
to $dm/dt$; optical emission will lie on the Rayleigh-Jeans tail of the distribution and thus scale as
$(dm/dt)^{1/4}$ \cite{Strubbe:2009qs,Lodato:2010xs}.  Although the connection between observed
monochromatic light curves and the fallback accretion rate is highly nontrivial, this accretion rate constitutes
the primary input for more sophisticated calculations of TDE emission.

Our preliminary survey of relativistic tidal disruption suggests that the spin magnitude $a/M$, inclination
$\iota$, and the polar angle at pericenter $\theta_p$ can all affect the fallback accretion rate provided $r_p$
is sufficiently small.  The newly discovered optical transient PS1-10jh \cite{Gezari:2012sa} may correspond
to a TDE with $r_p = 12M$ \cite{Lodato:2012} where relativistic effects can be significant as seen in
Fig.~\ref{F:r12M7E}.  The light curve associated with this event shows systematic differences with the
LKP09 model during the early rise that may be sensitive to relativistic corrections; a detailed comparison
between our model and this light curve is an important subject of future work.

We find that as $r_p$ approaches the radius of the marginally bound circular orbit (below which the star will
be directly captured by the SBH), relativistic corrections can increase the peak accretion rate
$dm_{\rm peak}/dt$ and reduce the time $t_{\rm peak}$ at which this peak accretion occurs by a factor of two. 
For a fixed value of $r_p$, relativistic effects are largest for large SBH spins that are anti-aligned with the
star's initial orbital angular momentum. However, as explored in detail in Sec.~\ref{SS:fallback}, there is a
great deal of degeneracy between $a/M$, $\iota$, and $\theta_p$.  This degeneracy is not terribly surprising
given the crudeness of our model, as each of these parameters can only increase or decrease the
distribution of fallback times.  Multi-frequency observations may be able to partially break this degeneracy.
Comparison of Figs.~\ref{F:r6M8E} and \ref{F:r6M8I} reveals a degeneracy between the fallback accretion
rates for TDEs of stars initially on inclined orbits and those on equatorial orbits of SBHs with lower spins.
Although the fallback accretion rates may be the same, the disks that subsequently form about the more
highly spinning SBHs should have smaller radii and thus higher temperatures.  Emission from these disks
should therefore be harder than that from the disks of SBHs with lower spins.  A more quantitative study of
the degeneracy between relativistic parameters is another subject of potential future work.

Our model of tidal disruption, like that of LKP09, is an example of what Guillochon and Ramirez-Ruiz
\cite{Guillochon:2012uc} describe as a ``freezing" model in that the distribution of the orbital constants of the
tidal debris is``frozen in" at pericenter.  This failure to account for the redistribution of orbital constants during
the small but nonzero time over which tidal disruption occurs can lead to several discrepancies in the
fallback accretion rate as discussed extensively in Guillochon and Ramirez-Ruiz \cite{Guillochon:2012uc}.
For fully-disruptive encounters like those considered in this paper, hydrodynamical simulations performed by
Guillochon and Ramirez-Ruiz suggest that $dm/dt$ is insensitive to $r_p$ when it is much below the tidal
radius $r_{\rm tid}$, in contrast to the predictions of freezing models.  The star's self-gravity is insufficient to
hold it together at pericenter for $r_p \ll r_{\rm tid}$, so it is inappropriate to apply freezing models to the
unperturbed star at $r_p$ in this regime.  Future work could explore applying relativistic freezing models to
the star at $r_{\rm tid}$ instead of $r_p$ as suggested by Guillochon and Ramirez-Ruiz; we would expect relativistic corrections to $dm/dt$ to be much
smaller in this case.

Another problem of freezing models is that they fail to account for pressure gradients that develop when the
star is tidally compressed \cite{Lodato:2008fr,Guillochon:2008gk}.  These pressure gradients can redistribute
material to more tightly bound orbits, leading to an increased feeding rate at early times.  Our study suggests
that this increase in the early rise could be misinterpreted as a relativistic effect if $r_p$ is small.  We hope to
go beyond the limitations of freezing models in future work on relativistic tidal disruption.  One subject
deserving special scrutiny is the conditions under which the tidal debris viscously evolves into an accretion
disk about the SBH.  Lense-Thirring precession \cite{Lense:1918zz} should delay this process for stars on
inclined orbits, helping to further break some of the degeneracy between spin magnitude $a/M$ and
inclination $\iota$.

Despite this need for additional work, the future prospects for using TDEs to measure SBH spins seem
highly promising.  The number of observed TDE candidates has increased dramatically in the past few years,
and future transient surveys should provide hundreds if not thousands of additional candidates
\cite{vanVelzen:2010jp}.  A portion of these candidates, like PS1-10jh \cite{Gezari:2012sa}, should have
sufficiently small $r_p$ such that relativistic effects will be significant.  We look forward to challenging our
model and its future refinements with this rich observational bounty.

\vspace{1cm}

{\bf Acknowledgements.} We would like to thank Glennys Farrar, Suvi Gezari, James Guillochon, Giuseppe
Lodato, Gabe Perez-Giz, Sterl Phinney, Enrico Ramirez-Ruiz, Roman Shcherbakov, Nick Stone, and Linda
Strubbe for useful conversations.

\appendix

\section{Kerr metric} \label{A:Kerr}

In this appendix, we provide explicit formulae for quantities relevant to describing tidal disruption in the Kerr
spacetime \cite{Kerr:1963ud}.  In Boyer-Lindquist coordinates \cite{Boyer:1966qh} and units where $G
= c = 1$, the Kerr metric takes the form
\begin{eqnarray} \label{E:met}
ds^2 &=& g_{\alpha\beta} dx^\alpha dx^\beta \nonumber \\
&=& -\left( 1 - \frac{2Mr}{\Sigma} \right) dt^2 - \frac{4Mar \sin^2 \theta}{\Sigma} dt d\phi \nonumber \\ 
&& + \frac{\Sigma}{\Delta} dr^2 + \Sigma~d\theta^2 + \frac{A}{\Sigma} \sin^2 \theta~d\phi^2
\end{eqnarray}
where $M$ is the mass of the black hole, $a/M$ is its dimensionless spin, and 
\begin{subequations} \label{E:metfun}
\begin{eqnarray}
\label{E:Sigma}
\Sigma &\equiv& r^2 + a^2 \cos^2 \theta~, \\
\label{E:Delta}
\Delta &\equiv& r^2 + a^2 - 2Mr~, \\
\label{E:A}
A &\equiv& (r^2 + a^2)^2 - \Delta a^2 \sin^2 \theta~.
\end{eqnarray}
\end{subequations}
The inverse of this metric is given by
\begin{equation}
g^{\alpha\beta} = 
\left(
\begin{matrix}
-\frac{A}{\Sigma\Delta} & 0 & 0 & -\frac{2Mar}{\Sigma\Delta} \\
0 & \frac{\Delta}{\Sigma} & 0 & 0 \\
0 & 0 & \frac{1}{\Sigma} & 0 \\
-\frac{2Mar}{\Sigma\Delta} & 0 & 0 & \frac{\Delta - a^2 \sin^2 \theta}{\Sigma\Delta\sin^2 \theta} 
\end{matrix}
\right)
\end{equation}
Massive particles move on timelike geodesics of the Kerr metric with 4-velocity
\begin{equation} \label{E:4v}
u^\alpha = (\dot{t}~\dot{r}~\dot{\theta}~\dot{\phi}),
\end{equation}
where overdots denote derivatives with respect to proper time $\tau$.  The Kerr metric possesses timelike
and azimuthal Killing vector fields, which we will denote by $(\partial/\partial t)^\alpha$ and
$(\partial/\partial \phi)^\alpha$ following the notation of Wald \cite{Wald:1984rg}.  This implies the existence
of a conserved specific energy and angular momentum 
\begin{subequations} \label{E:ELz}
\begin{eqnarray}
\label{E:Edef}
E &\equiv& -g_{\alpha\beta} u^\alpha \left( \frac{\partial}{\partial t} \right)^\beta~, \\
\label{E:Lzdef}
L_z &\equiv& g_{\alpha\beta} u^\alpha \left( \frac{\partial}{\partial \phi} \right)^\beta~.
\end{eqnarray}
\end{subequations}
The Kerr metric also possesses a Killing tensor \cite{Wald:1984rg,Walker:1970un}
\begin{equation} \label{E:Killtens}
K_{\alpha\beta} = \Sigma (l_{\alpha} n_{\beta} + n_{\alpha} l_{\beta})+ r^2 g_{\alpha\beta}~,
\end{equation}
where
\begin{subequations} \label{E:nullV}
\begin{eqnarray}
\label{E:ldef}
l^\alpha &\equiv& \frac{r^2 + a^2}{\Delta} \left( \frac{\partial}{\partial t} \right)^\alpha
+ \frac{a}{\Delta} \left( \frac{\partial}{\partial \phi} \right)^\alpha + \left( \frac{\partial}{\partial r} \right)^\alpha \\
\label{E:ndef}
n^\alpha &\equiv& \frac{r^2 + a^2}{2\Sigma} \left( \frac{\partial}{\partial t} \right)^\alpha
+ \frac{a}{2\Sigma} \left( \frac{\partial}{\partial \phi} \right)^\alpha - \frac{\Delta}{2\Sigma}
\left( \frac{\partial}{\partial r} \right)^\alpha \quad \quad
\end{eqnarray}
\end{subequations}
are the repeated principal null vectors found in a Petrov classification of the Weyl tensor
 \cite{Wald:1984rg,Petrov:1954,Petrov:1969}.  This implies the existence of a third constant of motion
\begin{eqnarray} \label{E:Qdef}
K &\equiv& K_{\alpha\beta} u^\alpha u^\beta \nonumber \\
&\equiv& Q + (L_z - aE)^2~.
\end{eqnarray}
The Carter constant $Q$ is defined in this way so that in the Newtonian limit ($r \to \infty$, $E \to 1$),
$Q \to L_{x}^2 + L_{y}^2$, where $L_x$ and $L_y$ are the $x$ and $y$ components of the
orbital angular momentum.  We can also define the inclination
\begin{equation} \label{E:incdef}
\cos \iota \equiv \frac{L_z}{\sqrt{Q + L_{z}^2}}
\end{equation}
which in the Newtonian limit converges to the angle between the SBH spin and angular momentum
${\mathbf L}_N$.

Eqs.~(\ref{E:ELz}) and (\ref{E:Qdef}), along with the normalization of the
4-velocity $g_{\alpha\beta} u^\alpha u^\beta = -1$, can be rearranged to provide equations of motion on
Kerr geodesics \cite{Carter:1968rr}
\begin{subequations} \label{E:EOM}
\begin{eqnarray}
\Sigma \dot{t} &=& \frac{AE- 2Mar L_z}{\Delta} \\
\label{E:dtdtau}
(\Sigma\dot{r})^2 &=& [E(r^2 + a^2) - aL_z]^2 \nonumber \\
\label{E:drdt}
&& -\Delta[r^2 + (L_z - aE)^2 + Q] \\
\label{E:dthetadt}
(\Sigma \dot{\theta})^2 &=& Q - L_{z}^2 \cot^2 \theta - a^2 (1-E^2) \cos^2 \theta \quad \quad \\
\label{E:dphidt}
\Sigma \dot{\phi} &=& L_z \csc^2 \theta + \frac{2MarE- a^2 L_z}{\Delta}~.
\end{eqnarray}
\end{subequations}

Evaluating the covariant derivatives of tensors in Boyer-Lindquist coordinates requires the use of
Christoffel symbols $\Gamma^{\alpha}_{\beta\gamma}$ \cite{Wald:1984rg}
\begin{eqnarray} \label{E:tenderiv}
\nabla _a T^{b_1 \cdots b_k} \,_{c_1 \cdots c_l} &=& \partial_a T^{b_1 \cdots b_k} \,_{c_1 \cdots c_l}
\nonumber \\
&& + \sum_i \Gamma^{b_i}_{ad} T^{b_1 \cdots d \cdots b_k} \,_{c_1 \cdots c_l}
\nonumber \\
&& - \sum_j \Gamma^{d}_{ac_j} T^{b_1 \cdots b_k} \,_{c_1 \cdots d \cdots c_l}~. \quad
\end{eqnarray}
The Christoffel symbols are symmetric under exchange of their lower two indices.  In Boyer-Lindquist
coordinates, the non-vanishing Christoffel symbols are
\begin{subequations} \label{E:Christ}
\begin{eqnarray}
\label{E:Crtt}
\Gamma^{r}_{tt} &\equiv& \frac{M\Delta}{\Sigma^3} (2r^2 - \Sigma) \\
\label{E:Cthtt}
\Gamma^{\theta}_{tt} &\equiv& -\frac{2Ma^2 r\sin \theta \cos \theta}{\Sigma^3} \\
\label{E:Crrr}
\Gamma^{r}_{rr} &\equiv& \frac{r}{\Sigma} - \frac{r-M}{\Delta} \\
\label{E:Cthrr}
\Gamma^{\theta}_{rr} &\equiv& \frac{a^2 \sin \theta \cos \theta}{\Sigma\Delta}  \\
\label{E:Crthth}
\Gamma^{r}_{\theta\theta} &\equiv& -\frac{r\Delta}{\Sigma} \\
\label{E:Cththth}
\Gamma^{\theta}_{\theta\theta} &\equiv& -\frac{a^2 \sin \theta \cos \theta}{\Sigma} \\
\label{E:Crpp}
\Gamma^{r}_{\phi\phi} &\equiv& -\frac{\Delta \sin^2 \theta}{\Sigma} \left[ r -
\frac{Ma^2 \sin^2 \theta}{\Sigma^2} (2r^2 - \Sigma) \right] \\
\label{E:Cthpp}
\Gamma^{\theta}_{\phi\phi} &\equiv& -\frac{\sin \theta \cos \theta}{\Sigma^3} [(r^2 + a^2)A - \Sigma\Delta
a^2 \sin^2 \theta] \quad \quad \\
\label{E:Cttr}
\Gamma^{t}_{tr} &\equiv& \frac{M(r^2 + a^2)}{\Sigma^2\Delta} (2r^2 - \Sigma) \\
\label{E:Cptr}
\Gamma^{\phi}_{tr} &\equiv& \frac{Ma}{\Sigma^2\Delta} (2r^2 - \Sigma) \\
\label{E:Cttth}
\Gamma^{t}_{t\theta} &\equiv& -\frac{2Ma^2 r\sin \theta \cos \theta}{\Sigma^2} \\
\label{E:Cptth}
\Gamma^{\phi}_{t\theta} &\equiv& -\frac{2Mar \cos \theta}{\Sigma^2 \sin \theta} \\
\label{E:Crtp}
\Gamma^{r}_{t\phi} &\equiv& -\frac{Ma\Delta \sin^2 \theta}{\Sigma^3} (2r^2 - \Sigma) \\
\label{E:Cthtp}
\Gamma^{\theta}_{t\phi} &\equiv& \frac{2Mar(r^2 + a^2) \sin \theta \cos \theta}{\Sigma^3} \\
\label{E:Crrth}
\Gamma^{r}_{r\theta} &\equiv& -\frac{a^2 \sin \theta \cos \theta}{\Sigma} \\
\label{E:Cthrth}
\Gamma^{\theta}_{r\theta} &\equiv& \frac{r}{\Sigma} \\
\label{E:Ctrp}
\Gamma^{t}_{r\phi} &\equiv& -\frac{Ma\sin^2 \theta}{\Sigma\Delta} \left[ \frac{2r^2}{\Sigma}(r^2 + a^2)
+ r^2 - a^2 \right] \\
\label{E:Cprp}
\Gamma^{\phi}_{r\phi} &\equiv& \frac{r}{\Sigma} - \frac{a^2\sin^2 \theta}{\Sigma\Delta} \left( r - M +
\frac{2Mr^2}{\Sigma} \right) \\
\label{E:Ctthp}
\Gamma^{t}_{\theta\phi} &\equiv& \frac{2Ma^3 r\sin^3 \theta \cos \theta}{\Sigma^2} \\
\label{E:Cpthp}
\Gamma^{\phi}_{\theta\phi} &\equiv& \frac{\cos \theta}{\sin \theta} \left( 1 +
\frac{2Ma^2r\sin^2 \theta}{\Sigma^2} \right)~.
\end{eqnarray}
\end{subequations}

\section{Fermi normal coordinates} \label{A:Fermi}

Although the Boyer-Lindquist coordinates utilized in the previous appendix are an excellent choice
for describing motion along geodesics of the Kerr spacetime, Fermi normal coordinates
\cite{Manasse:1963} provide a better description of the local neighborhood of a particle (such as
the center of mass of a tidally disrupting star) moving along such a geodesic.  These coordinates are
defined by first choosing a reference point $P_0$ on a central geodesic $G$ as the origin.  One then
chooses an orthonormal tetrad of 4-vectors $\boldsymbol{\lambda}_A$ in the tangent space at $P_0$,
where $\boldsymbol{\lambda}_0$ is the tangent vector to $G$ and $\boldsymbol{\lambda}_i$
$(i=1,2,3)$ are three spacelike vectors.  The point $(\tau, X^i)$ $(i=1,2,3)$ in Fermi normal coordinates
is reached by traveling a proper time $\tau$ along $G$ to a point $h(\tau)$, then traveling a proper
distance $s = [\sum_i (X^i)^2]^{1/2}$ along the orthogonal geodesic passing through
$h(\tau)$ whose tangent vector is $X^i \boldsymbol{\lambda}_i$.

Marck \cite{Marck:1983} constructed an explicit tetrad of 4-vectors $\boldsymbol{\lambda}_A$ that are
parallel propagated along timelike geodesics of the Kerr spacetime.  He provided components of these
4-vectors in the canonical symmetric orthornormal tetrad first introduced by Carter \cite{Carter:1968ks},
but to avoid introducing yet a third set of coordinates we provide these vectors in Boyer-Lindquist
coordinates.  $\lambda_{0}^\alpha$ is simply the 4-velocity of the center of mass, given explicitly by
Eqs.~(\ref{E:4v}) and (\ref{E:EOM}).  The Boyer-Lindquist components of the spacelike 4-vectors
$\lambda_{i}^\alpha$ are
\begin{subequations} \label{E:FNCv}
\begin{eqnarray}
\label{E:v1t}
\lambda_{1}^t &=& \frac{1}{K^{1/2}} \left[ \frac{\alpha (r^2 + a^2)r\dot{r}}{\Delta} + \beta a^2 \sin \theta
\cos \theta \dot{\theta} \right] \\
\label{E:v1r}
\lambda_{1}^r &=& \frac{\alpha r}{\Sigma K^{1/2}} [E(r^2 + a^2) - aL_z] \\
\label{E:v1th}
\lambda_{1}^\theta &=& \frac{\beta a\cos \theta}{\Sigma K^{1/2}} \left( aE\sin\theta - \frac{L_z}{\sin\theta}
\right) \\
\label{E:v1p}
\lambda_{1}^\phi &=& \frac{a}{K^{1/2}} \left( \frac{\alpha r\dot{r}}{\Delta}
+ \frac{\beta\cos\theta\dot{\theta}}{\sin\theta} \right) \\
\label{E:v2t}
\lambda_{2}^t &=& \frac{a}{K^{1/2}} \left[ \frac{(r^2 + a^2)\cos\theta\dot{r}}{\Delta} - r\sin\theta \dot{\theta}
\right] \\
\label{E:v2r}
\lambda_{2}^r &=& \frac{a\cos\theta}{\Sigma K^{1/2}} [E(r^2 + a^2) - aL_z] \\
\label{E:v2th}
\lambda_{2}^\theta &=& -\frac{r}{\Sigma K^{1/2}} \left( aE\sin\theta - \frac{L_z}{\sin\theta} \right) \\
\label{E:v2p}
\lambda_{2}^\phi &=& \frac{1}{K^{1/2}} \left( \frac{a^2\cos\theta\dot{r}}{\Delta}
- \frac{r\dot{\theta}}{\sin\theta} \right) \\
\label{E:v3t}
\lambda_{3}^t &=& \alpha\frac{r^2 + a^2}{\Sigma \Delta} [E(r^2 + a^2) - aL_z] \nonumber \\
&& - \beta\frac{a}{\Sigma} (aE\sin^2 \theta - L_z) \\
\label{E:v3r}
\lambda_{3}^r &=& \alpha \dot{r} \\
\label{E:v3th}
\lambda_{3}^\theta &=& \beta \dot{\theta} \\
\label{E:v3p}
\lambda_{3}^\phi &=& \frac{\alpha a}{\Sigma \Delta} [E(r^2 + a^2) - aL_z] - \frac{\beta}{\Sigma}
\left( aE - \frac{L_z}{\sin^2 \theta} \right) \quad \quad
\end{eqnarray}
\end{subequations}
where
\begin{subequations} \label{E:FNCcon}
\begin{eqnarray}
\label{E:alphacon}
\alpha &\equiv& \left( \frac{K - a^2 \cos^2 \theta}{r^2 + K} \right)^{1/2} \\
\beta &\equiv& \frac{1}{\alpha}~.
\end{eqnarray}
\end{subequations}

The basis 4-vectors $\boldsymbol{\lambda}_1$ and $\boldsymbol{\lambda}_3$ given above are equal
to $\boldsymbol{\tilde{\lambda}}_1$ and $\boldsymbol{\tilde{\lambda}}_3$ given in \cite{Marck:1983}, and
are not parallel propagated along the central geodesic $G$.  However, since we only make use of Fermi
normal coordinates at a single point along $G$ (the pericenter of the tidally disrupted star's orbit), we are
justified in using these vectors in our tetrad.  If the origin $P_0$ (the center of mass of a star) is located at
$x_{0}^\alpha$ in Boyer-Lindquist coordinates, an event located at $(0, X^i)$ in Fermi normal coordinates
will be located at
\begin{equation} \label{E:newx}
x^\alpha = x_{0}^\alpha + X^i \lambda_{i}^\alpha
\end{equation}
in Boyer-Lindquist coordinates.

\end{document}